\documentclass[10pt]{iopart}
\usepackage{iopams,times,graphicx,subfigure}

\makeatletter

\newcounter{theorem}
\@addtoreset{theorem}{section}
\renewcommand\thetheorem{\arabic{section}.\arabic{theorem}}
\newenvironment{definition}{\par\medskip\noindent\begingroup{\bf Definition
             \stepcounter{theorem}\thetheorem.}\ \itshape
             \def\@currentlabel{\thetheorem}}{\endgroup\par\medskip}
\newenvironment{lemma}{\par\medskip\noindent\begingroup{\bf Lemma
             \stepcounter{theorem}\thetheorem.}\ \itshape
             \def\@currentlabel{\thetheorem}}{\endgroup\par\medskip}
\newenvironment{theorem}{\par\medskip\noindent\begingroup{\bf Theorem
             \stepcounter{theorem}\thetheorem.}\ \itshape
             \def\@currentlabel{\thetheorem}}{\endgroup\par\medskip}

\newenvironment{proof}{\par\noindent{\bf Proof.} }{\proofbox\par\medskip}

\def\proofbox{\hfill{\ensuremath\Box}}

\usepackage{ulem}

\eqnobysec
\newcommand\partialderiv[3][]{\frac{\partial^{#1}#2}{\partial {#3}^{#1}}}
\def\eqref#1{(\ref{#1})}
\let\true@epsilon=\epsilon
\let\epsilon=\varepsilon
\def\Nm{N_-}
\def\Np{N_+}
\def\N{N}
\def\M{M}
\def\e{{\rm e}}
\def\r{r}
\def\s{s}

\def\journal#1&#2,{\begingroup \let\journal=\dummyjournal
               \it #1\unskip~\bf\ignorespaces #2\rm,\endgroup}
\def\dummyjournal{\errmessage{Reference foul up: nested \journal macros}}

\makeatother

\begin{document}
\title[Resonance and web structure in discrete soliton systems: the 2D Toda lattice]
  {Resonance and web structure in discrete soliton systems:\\the two-dimensional Toda lattice
  and its fully discrete and ultra-discrete analogues}
%\title[Soliton resonance and web structure in the 2D Toda lattice and its discrete versions]
%  {Soliton resonance and web structure in the two-dimensional Toda lattice
%  and its fully discrete and ultra-discrete versions}
\author{Ken-ichi Maruno$^1$\footnote[1]{E-mail: maruno@math.kyushu-u.ac.jp.} 
 and Gino Biondini$^2$}
\address{$^1$~Faculty of Mathematics, Kyushu University,
  Hakozaki, Higashi-ku, Fukuoka, 812-8581, Japan}
\address{$^2$~Department of Mathematics, State University of New York,
  Buffalo, NY 14260-2900, USA}
\date{\today}
\def\submitto#1{\vspace{28pt plus 10pt minus 18pt}
     \noindent{\small\rm Appeared to: {\it #1}\unskip, A/181581/PAP/46078\par}}

\begin{abstract}
We present a class of solutions of the two-dimensional Toda lattice equation, 
its fully discrete analogue and its ultra-discrete limit.
These solutions demonstrate the existence of soliton resonance and web-like 
structure in discrete integrable systems such as 
differential-difference equations, difference equations
and cellular automata (ultra-discrete equations).
\par
\kern\bigskipamount\noindent
\today
\end{abstract}

\kern-\bigskipamount
%{\bf Keywords}: Soliton Cellular Automaton, Soliton Resonance

\submitto{\JPA}

%%%%%%%%%%%%%%%%%%%%%%%%%%%%%%%%%%%%%%%%%%%%%%%%%%%%%%%%%%%%%%%%%%%%%%%%%
\section{Introduction}

The discretization of integrable systems is an important issue
in mathematical physics.
The most common situation is that in which some or all of the 
independent variables are discretized.
A discretization process in which the dependent variables are also 
discretized in addition to the
independent variables is known as ``ultra-discretization''. 
One of the most important ultra-discrete soliton systems is the 
so-called ``soliton cellular automaton'' 
(SCA)~\cite{Park,Takahashi2,Takahashi}.
A general method to obtain the SCA from discrete soliton equations
was proposed in Refs.~\cite{Junta,Tokihiro}
and involves using an appropriate limiting procedure.
Another issue which has received renewed interest in recent years is 
the phenomenon of soliton resonance,
which was first discovered for the 
Kadomtsev-Petviashvili (KP) equation~\cite{JFM1977v79p171}
(see also Refs.~\cite{LMP2002v62p91,PRL38p377}). 
More general resonant solutions possessing a web-like structure 
have recently been observed in a coupled KP (cKP)
system~\cite{JPhysA2002v35p6893,JPhysA2003v36p9533}
and for the KP equation itself~\cite{JPhysA2003v36p10519}.
In particular, the Wronskian formalism was used 
in Ref.~\cite{JPhysA2003v36p10519} to classify a class of 
resonant solutions of KP which also satisfy the Toda lattice hierarchy.
It was also conjectured in Ref.~\cite{JPhysA2003v36p10519} that 
resonance and web structure are not limited to KP and cKP, but 
rather they are a generic feature of integrable systems 
whose solutions can be expressed in terms of Wronskians.

The aim of this paper is to study 
soliton resonance and web structure
in discrete soliton systems.  
In particular, by studying a class of soliton solutions 
of the two-dimensional Toda lattice (2DTL) equation, of its fully 
discrete version, and of their ultra-discrete analogue 
which was recently introduced by Nagai et al~\cite{Mori,Nagai},
we show that an analogue to the class of 
solutions studied in Ref.~\cite{JPhysA2003v36p10519} can be defined 
for all three of these systems, and that a similar type 
of resonant solutions with web-like structure is produced as a result 
in all three of these systems.
To our knowledge, this is the first time that resonant behavior and
web structure are observed in discrete soliton systems.
These results also confirm that
soliton resonance and web-like structure are general features of 
two-dimensional integrable systems whose solutions can be 
expressed via the determinant formalism.

%%%%%%%%%%%%%%%%%%%%%%%%%%%%%%%%%%%%%%%%%%%%%%%%%%%%%%%%%%%%%%%%%%%%%%%%%
\section{The two-dimensional Toda lattice equation}

We start by considering
the two-dimensional Toda lattice (2DTL) equation,
\begin{equation}
\frac{\partial^2}{\partial x \partial t}Q_n
  =V_{n+1}-2V_n+V_{n-1}\,,
\label{2dTL}\\
\end{equation}
with $Q_n(x,t)=\log [1+V_n(x,t)]$.
Equation~\eqref{2dTL} can be written in bilinear form
\begin{equation}
\frac{\partial^2\tau_{n}}{\partial x \partial t}\,\,\tau_{n}
-\frac{\partial \tau_{n}}{\partial t}\,\,
\frac{\partial \tau_{n}}{\partial x}
=\tau_{n+1}\,\tau_{n-1}-\tau_{n}^2\,
\label{bilinear2dTL}
\end{equation}
through the transformation
\begin{equation}
V_n(x,t)=\frac{\partial^2}{\partial x \partial t}
\log \tau_{n}(x,t)\,.
\label{2dTLsoln}
\end{equation}
It is well-known that some solutions of the 2DTL equation can be written
via the
Casorati determinant form $\tau_{n}=\tau^{(\N)}_{n}$~\cite{PTP94p59}, with
\begin{equation}
\tau^{(\N)}_{n}= 
%\mathop{\rm Wr}\,(f^{(1)},\dots,f^{(\N)}):=
\left|\begin{array}{ccc}
f^{(1)}_{n} &\cdots &f^{(1)}_{n+\N-1}\\
\vdots &\ddots &\vdots\\
f^{(\N)}_{n} &\cdots &f^{(\N)}_{n+\N-1}
\end{array}\right|,
\label{2dTL:tau}
\end{equation}
where $\{f_n^{(1)}(x,t),\dots,f_n^{(\N)}(x,t)\}$ is a set of 
$\N$ linearly independent solutions of the linear equations
%where $\{f^{(i)}_n(x,t)\,|~i=1,\dots,\N\}$ is a
%linearly independent set of $\N$ solutions of the equations,
\begin{equation}
\frac{\partial f_{n}^{(i)}}{\partial x}= f_{n+1} \,,
\qquad
\frac{\partial f_{n}^{(i)}}{\partial t}= -f_{n-1}\,, 
%\label{2dTL:linear}
\nonumber
\end{equation}
for $1\le i\le \N$.
[Note that the superscript ``$(i)$'' does not denote differentiation here.]
For example, a two-soliton solution of the 2DTL is obtained by the set
$\{f^{(1)},f^{(2)}\}$, with
\begin{equation}
f^{(i)}_{n}= \e^{\theta^{({2i-1})}_{n}}+\e^{\theta^{({2i})}_{n}}\,, \qquad i=1,2\,,
\label{f2solitons}
\end{equation}
where the phases $\theta^{(j)}$ are given by linear functions of $(n,x,t)$:
\begin{equation}
\theta^{(j)}_n(x,t)=n\log p_j+ p_j x
-\frac{1}{p_j}t+ \theta^{(j)}_0\,, \qquad j=1,\dots,4\,,
\label{theta}
\end{equation}
with $p_1<p_2<p_3<p_4$. 
Equation~\eqref{f2solitons} can be extended to the
$\N$-soliton solution by considering $\{f^{(1)},\cdots,f^{(\N)}\}$,
with each $f^{(i)}$ defined according to Eq.~\eqref{f2solitons}.

On the other hand, solutions of the 2DTL equation
can also be obtained by the set of $\tau$~functions
$\{\tau^{(\N)}_{n}\,|~\N=1,\dots,\M\}$ with the choice of $f$-functions,
\begin{eqnarray}
f^{(i)}_{n}= f_{n+i-1}\,,\qquad 1< i\le \N\le \M\,,
\label{ftoda}
\\
\noalign{\noindent with}
f_{n}= \sum_{j=1}^\M \e^{\theta^{(j)}_{n}}\,.
\label{f0toda}
\end{eqnarray}
and with the phases $\theta^{(j)}_{n}$, $1\le i\le \M$ still given 
by Eq.~\eqref{theta}.
(Note that the meaning of $N$ and $M$ is the opposite of Ref.~\cite{JPhysA2003v36p10519}.)
If the $f$-functions are chosen according to eq.~\eqref{ftoda}, the
$\tau$~function $\tau_n^{(\N)}$ is then given by the Hankel determinant
\begin{equation}
\tau^{(\N)}_{n}=
\left|\begin{array}{ccc}
f_{n} &\cdots &f_{n+\N-1}\\
\vdots &\ddots &\vdots\\
f_{n+\N-1} &\cdots &f_{n+2\N-2}
\end{array}\right|,
\label{2dTL:todatau}
\end{equation}
for $1\le \N\le \M$.
It should be noted that, even when the set of functions 
$\{f_n^{(i)}\}_{i=1}^\N$ is chosen according to Eq.~\eqref{ftoda},
%one has
%\begin{equation}
%f^{(i)}_n= \partialderiv[i]{f_n}n,
%\end{equation}
%for $j=1,\dots,\N$, 
%and the Hankel determinant~\eqref{2dTL:todatau} becomes a Wronskian
%in the same sense as for the KP equation.
%Indeed, 
no derivatives appear in the $\tau$~function, and therefore 
Eq.~\eqref{2dTL:todatau} cannot be considered a Wronskian in the
same sense as for the KP equation 
(cf.\ Eq.~(1.9) in Ref.~\cite{JPhysA2003v36p10519}).
Nonetheless, this choice produces a similar outcome as in the KP 
equation.
Indeed, similarly to Ref.~\cite{JPhysA2003v36p10519}, we have the following: 

\begin{lemma}
\label{2dTL:tauN}
Let $f_{n}$ be given by Eq.~\eqref{f0toda}, with $\theta^{(j)}_n$
($j=1,\dots,\M$) given by Eq.~\eqref{theta}.
%\[
%f_{n}=\sum_{j=1}^\M \e^{\theta^{(j)}_{n}}\,,\qquad\mathrm{with}\quad
%\theta^{(j)}_{n}=n\log p_j+ p_j x
%-\frac{1}{p_j}t+ \theta^{(j)}_0\,.)
%\]
Then, for $1 \le \N \le \M-1$ 
the $\tau$~function defined by the Hankel determinant~\eqref{2dTL:todatau}
has the form
\begin{equation}
\label{2dTL:tauplus}
\displaystyle{
\tau^{(\N)}_{n}=
  \kern-1em \sum_{1\le i_1<\cdots<i_\N\le \M}
    \Delta(i_1,\dots,i_\N)\,\,
\exp\bigg(\sum_{j=1}^\N\theta^{(i_j)}_{n}\bigg)\,,}
\label{tauplus}
\end{equation}
where $\Delta(i_1,\dots,i_\N)$ is the square 
of the van der Monde determinant,
\[
\Delta(i_1,\dots,i_\N)=\prod_{1\le j<l\le \N} 
(p_{i_j}-p_{i_l})^2\,.
\]
\end{lemma}
\begin{proof}
Apply the Binet-Cauchy theorem to Eq.~\eqref{2dTL:todatau}, 
as in Ref.~\cite{JPhysA2003v36p10519}.
\end{proof}

An immediate consequence of Lemma~\ref{2dTL:tauN}
is that the $\tau$~function $\tau^{(\N)}_{n}$ is 
positive definite, and therefore 
all the solutions generated by it are non-singular.
Like its analogue in the KP equation \cite{JPhysA2003v36p10519},
the above $\tau$~function produces soliton solutions 
of resonant type with web structure.
More precisely, in the next section we show that, like its analogue in the
KP equation, the above $\tau$~function produces an $(\Np,\Nm)$-soliton 
solution, that is, a solution with $\Np=\M-\N$ asymptotic line solitons
as $n\to \infty$ and $\Nm=\N$ asymptotic line solitons as $n\to-\infty.$

Before we turn our attention to resonant solutions, however, it is useful 
to take a look at the one-soliton solution of the 2DTL equation.
Let us introduce the function
\begin{equation}
w_n(x,t)= \partialderiv{}x \log \tau_n(x,t)\,,
\label{e:wndef}
\end{equation}
so that the solution of the 2DTL equation is given by
\begin{equation}
V_n(x,t)= \partialderiv{}t w_n(x,t)\,.
\end{equation}
If $\tau_n=e^{\theta_n^{(1)}}+e^{\theta_n^{(2)}}$, 
with $\theta_n^{(i)}$ given by Eq.~\eqref{theta} and $p_1<p_2$, 
then $w_n$ is given by
\[\begin{array}{lllll}
w_n&=&{\frac{1}{2}(p_1+p_2)+\frac{1}{2}(p_1-p_2){\rm
tanh}\frac{1}{2}(\theta_n^{(1)}-\theta_n^{(2)}) }\\[1.2ex]
&\longrightarrow&
\left\{
\begin{array}{ll}
p_1 &\quad{\rm as}~~x\to\infty,\\
p_2 &\quad{\rm as}~~x\to -\infty,
\end{array}
\right. 
\end{array}
\]
which leads to the one-soliton solution of the 2DTL equation:
\begin{equation}
\label{e:1-soliton}
V_n= %\frac{\partial^2}{\partial t \partial x} \log\tau_n=
-\frac{1}{4}(p_1-p_2)\left(\frac{1}{p_1}-\frac{1}{p_2}\right)
{\rm sech}^2 \frac{1}{2}(\theta_n^{(1)}-\theta_n^{(2)})\,.
\end{equation}
In the $x$-$n$ plane, this solution describes a plane wave
$u=\Phi({\bf k}\cdot{\bf x}-\omega\,t)$ with ${\bf x}=(x,n)$, 
having wavenumber vector ${\bf k}=(k_x,k_n)$ and frequency $\omega$
given by:
\[
{\bf k}=(p_1-p_2,~\log p_1-\log p_2)=:{\bf k}_{1,2},\quad
\omega=\frac{1}{p_1}-\frac{1}{p_2}=:\omega_{1,2}.
\]
The soliton parameters $({\bf k},~\omega)$ satisfy the 
dispersion relation $\omega k_x+2\cosh k_n-2=0$.  
The above one-soliton solution~\eqref{e:1-soliton} is referred to as 
a line soliton, since in the $x$-$n$ plane it is localized around
the (contour) line $\theta_n^{(1)}=\theta_n^{(2)}$.
Since in this paper we are interested in the pattern of soliton solutions
in the $x$-$n$ plane, we will refer to $c=dx/dn$ as the \textit{velocity} 
of the line soliton in the $x$ direction.
(That is, $c=0$ indicates the direction of the positive $n$-axis.)
For the soliton solution in Eq.~\eqref{e:1-soliton}, 
this velocity is~$c_{1,2}$,
where 
\begin{equation}
c_{i,j}=-(\log p_i-\log p_j)/(p_i-p_j)\,.
\label{e:solitonvelocity}
\end{equation}

%%%%%%%%%%%%%%%%%%%%%%%%%%%%%%%%%%%%%%%%%%%%%%%%%%%%%%%%%%%%%%%%%%%%%%%%%
\section{Resonance and web structure in the two-dimensional Toda lattice equation}

We first consider $(\Np,1)$-soliton solutions, i.e., solutions obtained when
$\N=1$.
In particular, we start with (2,1)-soliton solutions
(i.e., $\N=1$ and $\M=3$), 
whose $\tau$-function is given by
\[
\tau_n= e^{\theta_n^{(1)}} + e^{\theta_n^{(2)}} + e^{\theta_n^{(3)}}\,.
\]
with $p_1<p_2<p_3$ without loss of generality.
The corresponding function $w_n$ describes the confluence of two shocks:
two shocks for $n\to \infty$ (each corresponding to a line soliton for $V_n$)
with velocities $c_{1,2}$ and $c_{2,3}$
merge into a single shock for $n\to-\infty$ with velocity $c_{1,3}$,
with $c_{i,j}$ given by Eq.~\eqref{e:solitonvelocity} in all cases.
%$c_{i,j}=-(\log p_i-\log p_j)/(p_i-p_j)$ in all cases.
%$c_{1,2}=-(\log p_1-\log p_2)/(p_1-p_2)$ and
%$c_{2,3}=-(\log p_2-\log p_3)/(p_2-p_3)$, 
%and the single shock for $n\to\infty$
%has velocity $c_{1,3}=-(\log p_1-\log p_3)/(p_1-p_3)$.
%
This Y-shape interaction represents a resonance of three line solitons.
The resonance conditions for three solitons with wavenumber vectors
$\{{\bf k}_{i,j}\,|~1\le i<j\le 3\}$ and frequencies
$\{\omega_{i,j}\,|~1\le i<j\le 3\}$
are given by
\begin{equation}
\label{resonancecondition}
{\bf k}_{1,2}+{\bf k}_{2,3}={\bf k}_{1,3},\quad {\rm and}\quad
\omega_{1,2}+\omega_{2,3}=\omega_{1,3},
\end{equation}
which are trivially satisfied by those line solitons.
We should point out that this solution is also the resonant case
of the ordinary 2-soliton solution of the 2DTL equation.
As mentioned earlier, ordinary 2-soliton solutions are given by the
$\N=2$ $\tau$-function~\eqref{2dTL:tau} with~\eqref{f2solitons}.
The explicit form of the $\tau_n^{(2)}$-function is
\[
\fl
\tau_n^{(2)}= (p_1-p_3)\,e^{\theta_n^{(1,3)}}
    + (p_1-p_4)\,e^{\theta_n^{(1,4)}}
    + (p_2-p_3)\,e^{\theta_n^{(2,3)}}
    + (p_2-p_4)\,e^{\theta_n^{(2,4)}},
\]
where for brevity we introduced the notation 
$\theta_n^{(i,j)}=\theta_n^{(i)}+\theta_n^{(j)}$, 
and where $\theta_n^{(i)}$ is given by Eq.~\eqref{theta}, as before.
%$\theta_n^{(i)}=n\log p_i+ p_i x-\frac{1}{p_i}t+ \theta^{(i)}_0$.
Note that if $p_2=p_3$, the $\tau_n^{(2)}$-function can be written as
\[
\fl
\tau_n^{(2)}= e^{\theta_n^{(1)}+\theta_n^{(2)}+\theta_n^{(4)}}\big[
(p_1-p_3)\,\Delta\,e^{-\theta_n^{(4)}} + (p_1-p_4)\,e^{-\theta_n^{(2)}}
+(p_2-p_4)\,e^{-\theta_n^{(1)}}\big]\,,
\]
where $\Delta=\exp(\theta_0^{(3)}-\theta_0^{(2)})=$\,constant.
Since the exponential factor 
$ e^{\theta_n^{(1)}+\theta_n^{(2)}+\theta_n^{(4)}} $ gives
zero contribution to the solution~
$V_n=\partial_{t}\partial_{x}\log\tau_n^{(2)}$,
the above $\tau_n^{(2)}$-function is equivalent to a
(2,1)-soliton solution except for the signs of the phases
(more precisely, it is a (1,2)-soliton).
% that is, the resonant solution with confluence of solitons.
Note also that the condition $p_2=p_3$ is nothing else but the 
resonance condition,
and it describes the limiting case of an infinite phase shift
in the ordinary 2-soliton solution,
where the phase shift between the solitons as $n\to\pm\infty$ is
given by
\begin{equation}
\delta= {(p_1-p_3)(p_2-p_4)}/[{(p_2-p_3)(p_1-p_4)}]\,.
\label{e:phaseshift}
\end{equation}
The resonance process for the $(\Np,1)$-soliton solutions of the 2DTL equation 
can be expressed as a generalization of the confluence of shocks discussed 
above (cf.~Ref.~\cite{JPhysA2003v36p10519}). 

We next consider more general $(\Np,\Nm)$-soliton solutions.  
Following Ref.~\cite{JPhysA2003v36p10519}, we can describe 
the asymptotic pattern of the solution in the general case $\N\ne1$
by introducing a local coordinate frame $(\xi,n)$ in order to study 
the asymptotics for large $|n|$, with
\[
x= c\,n + \xi\,.
\]
The phase functions $\theta_n^{(i)}$ in $f$ in Eq.~\eqref{theta} then become
\[
\theta_n^{(i)}
= p_i\xi + \eta_i(c)\,n + \theta_0^{(i)}\,, \quad{\rm for}\quad
i=1,\dots, \M\,,
\]
with
\[
\eta_i(c):=p_i(c+(1/p_i)\,\log p_i)\,.
\]
Without loss of generality, we assume an ordering for the
parameters $\{p_i\,|~i=1,\dots,\M\}$: 
$0<p_1<p_2<\cdots<p_\M\,$.
Then one can easily show that the lines $\eta=\eta_i(c)$ are in general
position; that is, each line $\eta=\eta_i(c)$ intersects with all other
lines at $\M-1$ distinct points in the $c$-$\eta$ plane;
in other words, only {\it two} lines meets at each intersection point.

The goal is now to find the dominant exponential terms in the
$\tau_n^{(\N)}$-function~\eqref{tauplus} 
for $n\to \pm\infty$ as a function of
the velocity $c$. First note that if only one exponential is dominant,
then
$w_n=\partial_x \log\tau_n^{(\N)}$ is just a constant,
and therefore the solution $V_n=\partial_t w_n$ is zero.
Then, nontrivial contributions to $V_n$
arise when one can find
{\it two} exponential terms which dominate over the others.
Note that because the intersections of the $\eta_i$'s are always pairwise,
three or more terms cannot make a dominant balance for large $|n|$.
In the case of $(\Np,1)$-soliton solutions, it is easy to see that
at each $c$ the dominant exponential term for $n\to-\infty$ is provided
by only $\eta_1$ and/or $\eta_\M$, and therefore there is only one shock
($\Nm=1$) moving with velocity~$c_{1,\M}$
%=-(\log p_\M-\log p_1)/(p_\M-p_1)$ 
corresponding to
the intersection point of $\eta_1$ and $\eta_\M$.
On the other hand, as $n\to \infty$, each term $\eta_j$ can become
dominant for some $c$,
and at each intersection point $\eta_j=\eta_{j+1}$
the two exponential terms corresponding to $\eta_j$ and $\eta_{j+1}$
give a dominant balance;
therefore there are $\Np=\M-1$ shocks moving with
velocities~$c_{j,j+1}$ %=-(\log p_{j+1}-\log p_j)/(p_{j+1}-p_j)$, 
for $j=1,\dots,\M-1$.

In the general case, $\N\ne1$, the $\tau_n^{(\N)}$-function
in~\eqref{tauplus}
involves exponential terms having combinations of phases.
In this case the exponential terms that make a dominant balance
can be found using the same methods as in Ref.~\cite{JPhysA2003v36p10519}.
Let us first define the {\it level of intersection} of the~$\eta_i(c)$.
Note that $c_{i,j}$ in Eq.~\eqref{e:solitonvelocity}
identifies the intersection point of $\eta_i(c)$ and $\eta_j(c)$,
i.e., $\eta_i(c_{i,j})=\eta_j(c_{i,j})$.

\begin{definition}
We define the level of intersection, denoted by $\sigma_{i,j}$,
as the number of other $\eta_l$'s that are smaller
than $\eta_i(c_{i,j})=\eta_j(c_{i,j})$ at $c=c_{i,j}$.
That is,
\[
\sigma_{i,j}:=\big|\{\eta_l\,|~\eta_l(c_{i,j})<\eta_i(c_{i,j})
=\eta_j(c_{i,j})\}\big|\,.
\]
We also define $I(s)$ as the set of pairs $(\eta_i,\eta_j)$ having
the level $\sigma_{i,j}=s$, namely
\[ 
I(s):=\{(\eta_i,\eta_j)\,|~\sigma_{i,j}=s,~{\rm for}~i<j\,\}.
\]
\end{definition}
\noindent
The level of intersection lies in the range
$0\le\sigma_{i,j}\le \M-2$.
Note also that the total number of pairs $(\eta_i,\eta_j)$ is
\[
\left(\!\!\!\begin{array}{cc}\M\\2\end{array}\!\!\!\right)=\frac{1}{2} \M(\M-1)=
             \sum_{n=0}^{\M-2}|I(s)|\,.
\]
One can show that:
\begin{lemma}
\label{l:intersectingpairs}
The set $I(s)$ is given by
\[
I(s)=\{(\eta_i,\eta_{\M-s+i-1})\,|~i=1,\dots,s+1\}\,.
\]
\end{lemma}
\begin{proof}
From the assumption $q_1<q_2<\cdots<q_\M$, we have the following 
inequality
at $c=c_{i,j}$~ (i.e. $\eta_i=\eta_j$) for $i<j$,
\[
\eta_1,\dots,\eta_{i-1},\,\eta_{j+1},\dots,\eta_\M
<\eta_i=\eta_{j}
<\eta_{i+1},\dots,\eta_{j-1}\,.
\]
Then taking $j=\M-s-1$ leads to the assertion of the Lemma.
\end{proof} 

\noindent
Now define $\Nm=\N$ and $\Np=\M-\N$.
The above lemma indicates that, for each intersecting pair $(\eta_i,\eta_j)$
with the level $\Np-1$ ($\Nm-1$),
there are $\Nm-1$ terms $\eta_l$'s which are larger (smaller)
than~$\eta_i=\eta_j$.
Then the sum of those $\Nm-1$ terms with either $\eta_i$ or $\eta_j$
provides two dominant exponents in the $\tau_n^{(\N)}$-function
for $n\to \infty$ $(n\to -\infty)$ (see more detail in the proof of
Theorem~\ref{t:asymptotic}).
Note also that $|I(N_\pm-1)|=N_\mp$.
Now we can state our main theorem:

\begin{theorem}
\label{t:asymptotic}
Let\, $w_n$\, be defined by Eq.~\eqref{e:wndef},
%\[
%w_n= \frac{\partial}{\partial x} \log \tau_n^{(\N)}\,,
%\]
with $\tau_n^{(\N)}$ given by Eq.~\eqref{tauplus}.
Then\, $w_n$\, has the following asymptotics for $n\to\pm\infty:$
\renewcommand\labelenumi{{\rm(\roman{enumi})}}
\begin{enumerate}
\itemsep 0pt
\parsep 0pt
\item
For\, $n\to \infty$\, and\, $x=c_{i,\Nm+i}\,n+\xi$ ~for $i=1,\dots,\Np$\,,
\[
\kern-\leftmargini
w_n~\longrightarrow~\left\{\!\!\begin{array}{ll}
K_i(+,+):= \sum_{j=i+1}^{\N+i} p_j &\mathrm{as}~~\xi\to\infty\,,\\
K_i(-,+):= \sum_{j=i}^{\N+i-1} p_j &\mathrm{as}~~\xi\to-\infty\,.
\end{array}\right.
\]
\item
For\, $n\to -\infty$\, and\, $x=c_{i,\Np+i}\,n+\xi$~ for $i=1,\dots,\Nm$\,,
\[
\kern-\leftmargini
w_n~\longrightarrow~\left\{\!\!\begin{array}{ll}
K_i(+,-):=  \sum_{j=1}^{i-1} p_j
             + \sum_{j=1}^{\N-i+1} p_{\M-j+1} &\mathrm{as}~~
\xi\to \infty\,,\\
K_i(-,-):= \sum_{j=1}^i p_j
             + \sum_{j=1}^{\N-1} p_{\M-j+i} &\mathrm{as}~~
\xi\to -\infty\,.
\end{array}\right.\kern-0.4em
\]
\end{enumerate}
where\, $c_{i,j}$ is given by Eq.~\eqref{e:solitonvelocity}.
%=-(\log p_{j}-\log p_i)/(p_{j}-p_i)$.
\end{theorem}
\begin{proof}
First note that at the point $\eta_i=\eta_{\N+i}$, i.e.,
$(\eta_i,\eta_{\N+i})\in I(\Np-1)$, from Lemma~\ref{l:intersectingpairs}
we have the inequality,
\[
\eta_i=\eta_{\N+i}<
\underbrace{\phantom{\bigg|}\kern-0.2em
\eta_{i+1},\eta_{i+2},\dots,\eta_{i+\N-1}}_{\N-1}\,.
\]
This implies that, for $c=-(\log p_{\N+i}-\log p_i)/(p_{\N+i}-p_i)$,
the following two exponential terms in the
$\tau_n^{(\N)}$-function in Lemma~\ref{2dTL:tauN},
\[
\exp\bigg(\sum_{j=i}^{\N+i-1} \theta_n^{(j)}\bigg)\,, 
\qquad%\quad\mathrm{and}\quad
\exp\bigg(\sum_{j=i+1}^{\N+i} \theta_n^{(j)}\bigg) \,,
\]
provide the dominant terms for $n\to \infty$.
Note that the condition $\eta_i=\eta_{\N+i}$ leads to
$c=c_{i,\N+i}=-(\log p_{\N+i}-\log p_i)/(p_{\N+i}-p_i)$.
Thus the function $w_n$ can be approximated by the following form along
$x=c_{i,\N+i}\,n+\xi$ for $n\to \infty$:
\[
\begin{array}{ll}
w_n\kern-0.4em &\sim~ \displaystyle \frac{\partial}{\partial \xi}\log\big(
\Delta_i(-,+)e^{K_i(-,+)\xi}+\Delta_i(+,+)e^{K_i(+,+)\xi} \big)
\\[1.6ex]
&=~ \displaystyle{
\frac{K_i(-,+)\Delta_i(-,+)e^{K_i(-,+)\xi}+K_i(+,+)
\Delta_i(+,+)e^{K_i(+,+)\xi}}
{\Delta_i(-,+)e^{K_i(-,-)\xi}+\Delta_i(+,+)e^{K_i(+,+)\xi}}\,,}
\\[1.6ex]
&=~ \displaystyle{
\frac{K_i(-,+)\Delta_i(-,+)e^{(p_i-p_{\N+i})\xi}+K_i(+,+)\Delta_i(+,+)}
             {\Delta_i(-,+)e^{ (p_i-p_{\N+i})\xi}+\Delta_i(+,+)}\,,}
\end{array}
\]
where
\begin{eqnarray*}
\Delta_i(-,+)= \Delta(i,\dots,\N+i-1)\,\,
             \exp\bigg(\sum_{j=i}^{\N+i-1} \theta_0^{(j)}\bigg)
\\
\Delta_i(+,+)= \Delta(i+1,\dots,\N+i)\,\,
             \exp\bigg(\sum_{j=i+1}^{\N+i} \theta_0^{(j)}\bigg)\,.
\end{eqnarray*}
Now, from $p_i<p_{\N+i}$~it is obvious that $w_n$ has the desired 
asymptotics
as $\xi\to\pm\infty$ for $n\to \infty$.

Similarly,
for the case of $(\eta_i,\eta_{\Np+i})\in I(\Nm-1)$ we have the inequality
\[
\underbrace{\phantom{\bigg|}\kern-0.2em
\eta_1,\eta_2,\dots,\eta_{i-1},\eta_{\Np+i+1}\dots,\eta_\M}_{\N-1}
<\eta_i=\eta_{\Np+i}\,.
\]
Then the dominant terms in the $\tau_n^{(\N)}$-function on 
$x=c_{i,\Np+i}n+\xi$ for
$n\to -\infty$ are given by the exponential terms
\begin{equation*}
\fl
\exp\bigg(\sum_{j=1}^i \,\theta_n^{(j)} ~
             + \sum_{j=1}^{\N-i} \,\theta_{n}^{(\M-j+1)}\bigg)\,,
\qquad %\quad {\rm and} \quad
\exp\bigg(\sum_{j=1}^{i-1} \,\theta_n^{(j)} ~
             + \sum_{j=1}^{\N-i+1} \,\theta_{n}^{(\M-j+1)}\bigg)\,.
\end{equation*}
Then, following the same argument as before, we obtain the desired asymptotics
as $\xi\to\pm \infty$ for $n\to -\infty$.

For other values of $c$, that is for $c\ne c_{i,\Nm+i}$
and $c\ne c_{i,\Np+i}$, just one exponential term is dominant,
and thus $w_n$ approaches a constant as $|n|\to\infty$.
This completes the proof.
%\break{$\phantom.$}
\end{proof}

%%%%%%%%%%%%%%%%%%%%%%%%%%%%%%%%%%%%%%%%%%%%%%%%%%%%%%%%%%%%%%%%%%%%%%%%%
\begin{figure}[t!]
\centerline{
\includegraphics[scale=0.675]{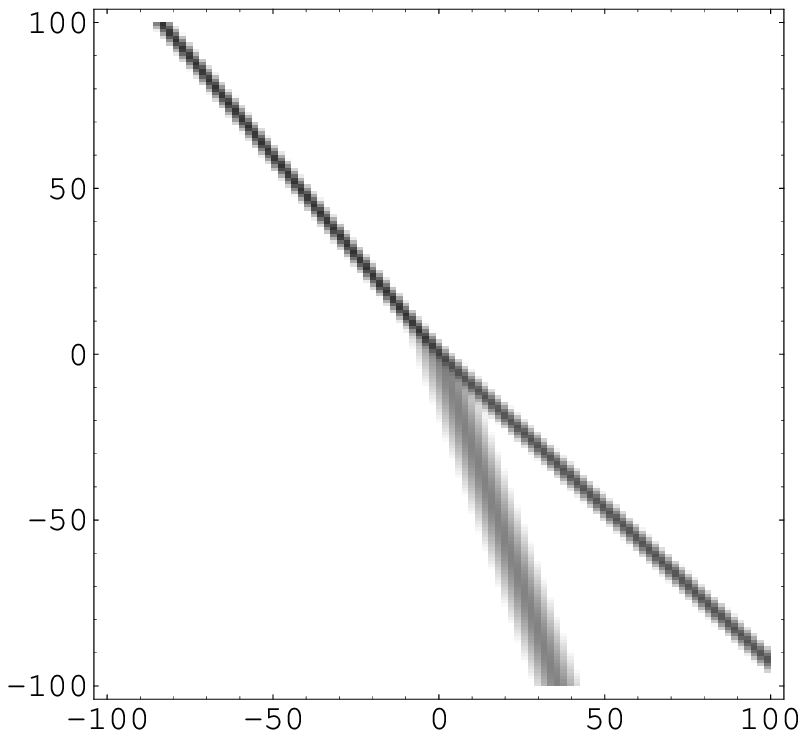}\quad
\includegraphics[scale=0.675]{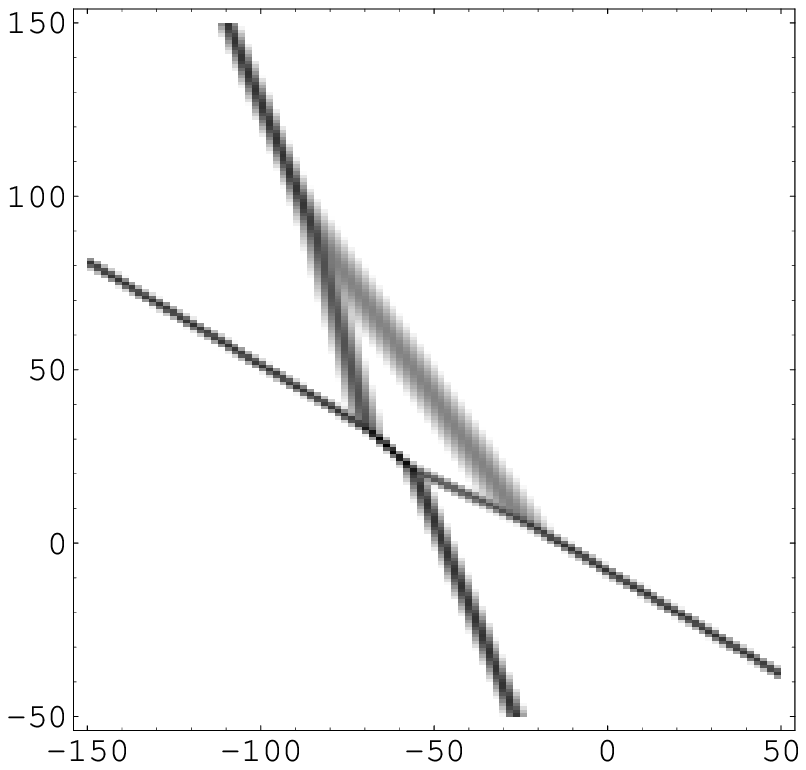}}
\kern-0.385\textwidth
\hbox to \textwidth{\hss(a)\kern0em\hss(b)\kern4em}
\kern+0.355\textwidth
\centerline{
\includegraphics[scale=0.675]{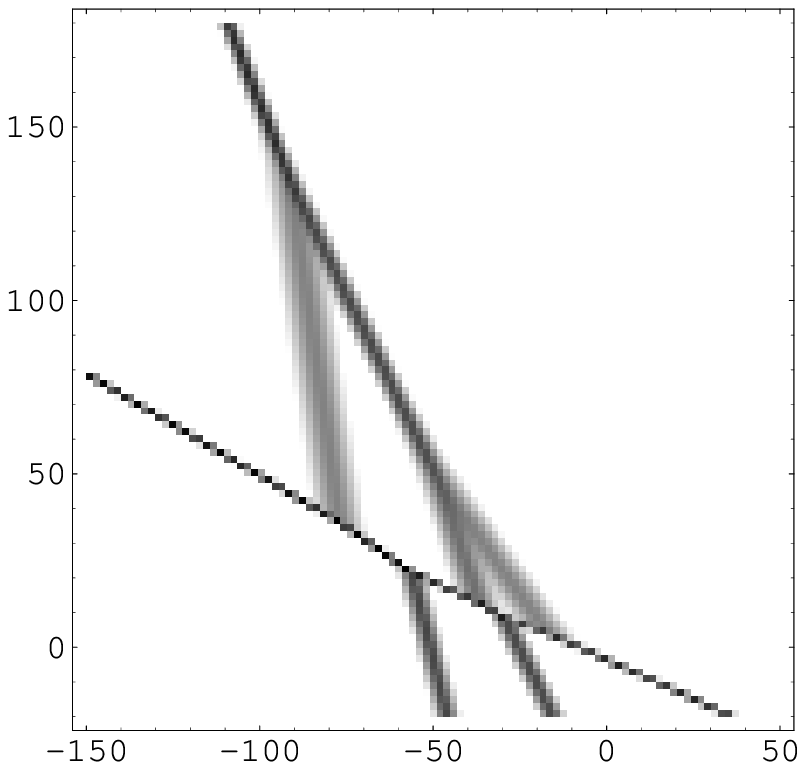}\quad
\includegraphics[scale=0.675]{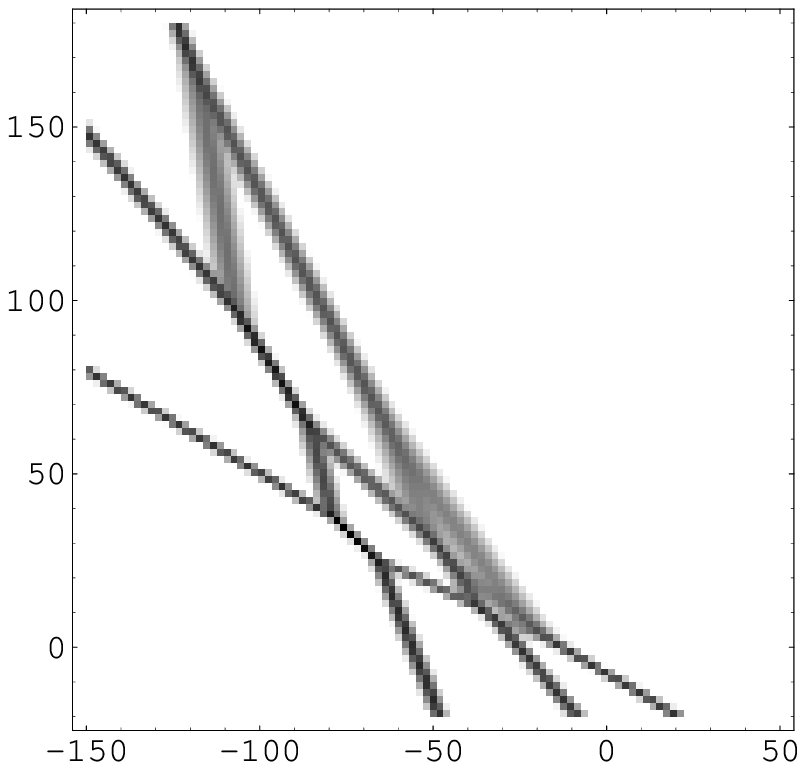}}
\kern-0.385\textwidth
\hbox to \textwidth{\hss(c)\kern0em\hss(d)\kern4em}
\kern+0.355\textwidth
\caption{Resonant solutions of the two-dimensional Toda lattice:
(a)~(2,1)-soliton solution (i.e., a Y-junction) at $t=0$, 
with $\N=1$, $\M=3$, $p_1=1/4$, $p_2=1/2$, $p_3=2$;
(b)~(2,2)-soliton solution at $t=14$, 
with $\N=2$, $\M=4$, $p_1=1/8$, $p_2=1/2$, $p_3=1$ and $p_4=4$;
(c)~(3,2)-soliton solution at $t=10$, 
with $\N=2$, $\M=5$, $p_1=1/10$, $p_2=1/5$, $p_3=1/2$, $p_4=1$ and
$p_5=6$;
(d)~(3,3)-soliton solution at $t=14$,
with $\N=3$, $\M=6$, $p_1=1/10$, $p_2=1/4$, $p_3=1/2$ and $p_4=1$, 
$p_5=2$, $p_6=4$.
In all cases the horizontal axis is $n$ and the vertical axis is $x$,
and each figure is a plot of $q(n,x,t)$ in logarithmic grayscale.
Note that the values of $n$ in the horizontal axis are discrete.}
\label{f:2dtoda}
\end{figure}
%%%%%%%%%%%%%%%%%%%%%%%%%%%%%%%%%%%%%%%%%%%%%%%%%%%%%%%%%%%%%%%%%%%%%%%%%

Theorem~\ref{t:asymptotic} determines the complete structure of 
asymptotic patterns of the solutions $V_n(x,t)$ given by~\eqref{2dTL:todatau}. 
Indeed, Theorem~\ref{t:asymptotic} can be summarized as follows:
As $n\to \infty$, the function $w_n$ has $\Np$ jumps, moving with velocities
$c_{j,\Nm+j}$ for $j=1,\dots,\Np$;
as $n\to -\infty$, $w_n$ has $\Nm$ jumps, moving with velocities
$c_{i,\Np+i}$ for $i=1,\dots,\Nm$.
Since each jump represents a line soliton for $V_n(x,t)$,
the whole solution therefore represents an $(\Np,\Nm)$-soliton.
The velocity of each of the asymptotic line solitons in the 
$(\Np,\Nm)$-soliton
is determined from the $c$-$\eta$ graph of the levels of intersections.
As an example, in Fig.~\ref{f:2dtoda} we show 
a (2,1)-soliton solution (also called a Y-shape solution,
or a Y-junction),
a (2,2)-soliton solution,
a (2,3)-soliton solution and a (3,3)-soliton solution. 

Note that, given a set of $\M$ phases (as determined by the parameters
$p_i$ for $i=1,\dots,\M$), the same graph can be used for
any $(\Np,\Nm)$-soliton with $\Nm+\Np=\M$.
In particular, if $\M=2\N$, we have $\Np=\Nm=\N$, 
and Theorem~\ref{t:asymptotic}
implies that the velocities of the $\N$ incoming solitons 
are equal to those
of the $\N$ outgoing solitons. 
%%%(Here, we call solitons at $n\to \infty$
%%%($n\to -\infty$) incoming(outgoing) solitons.)
In the case of the ordinary multi-soliton solution
of the 2DTL equation, the $\tau$-function~\eqref{2dTL:tau} does not
contain all the possible combinations of phases,
and therefore the theorem should be modified. 
However, the key idea of using the levels of intersection 
for the asymptotic analysis is still applicable. 
In fact, by considering the $\tau_n^{(\N)}$-function given by the
Casorati determinant (\ref{2dTL:tau}) 
with $f_{i}=e^{\theta_n^{(2i-1)}}+e^{\theta_n^{(2i)}}$ for
$i=1,\dots,\N$ and $p_1<p_2<\cdots<p_{2\N}$,
one can find the asymptotic velocities for the ordinary $\N$-soliton
solutions as 
$c_{2i-1,2i}=-(\log p_{2i}-\log p_{2i-1})/(p_{2i}-p_{2i-1})$ 
Note that these velocities are different from those of the resonant
$\N$-soliton solutions.

Note also that even when $\Np=\Nm=\N$, the interaction pattern of resonant 
soliton solutions differs from that of ordinary $\N$-soliton solutions.
As seen from Fig.~\ref{f:2dtoda}, the resonant solutions of the 2DTL obtained 
from Eq.~\eqref{2dTL:todatau} are very similar to the solitons of 
the KP and coupled KP equation~\cite{JPhysA2003v36p10519,JPhysA2002v35p6893,JPhysA2003v36p9533}, 
where such solutions were called ``spider-web'' solitons.
(In contrast, an ordinary $\N$-soliton solution produces a 
simple pattern of $\N$ intersecting lines.)
The web structure manifests itself in the number of bounded regions,
the number of vertices and the number of intermediate solitons,
which are respectively $(\Nm-1)(\Np-1)$,\ $2\Nm\Np-\M$\,\ and\,\ $3\Nm\Np-2\M$
for an $(\Np,\Nm)$-soliton solution~\cite{JPhysA2003v36p10519}.
(In contrast, an ordinary $\N$-soliton solution has $(\N-1)(\N-2)/2$ 
bounded regions and $\N(\N-1)/2$ interaction vertices.)\,\ 
Finally, it should be noted that, as in the KP equation, 
only the Y-shape solution is a traveling wave solution.  
All other resonant solutions (as well as ordinary 
$\N$-soliton solutions with $\N\ge3$) have a
time-dependent shape, as shown in Ref.~\cite{JPhysA2003v36p10519}.

%%%%%%%%%%%%%%%%%%%%%%%%%%%%%%%%%%%%%%%%%%%%%%%%%%%%%%%%%%%%%%%%%%%%%%%%%
\section{The fully discrete 2D Toda lattice equation}

The 2DTL equation~\eqref{2dTL} 
is a differential-difference evolution equation, 
since only one of the independent variables is discrete, 
while the other two are continuous. 
Hereafter, we refer to Eq.~\eqref{2dTL} as a semi-continuous case.
We now consider a fully discrete analogue of the 2DTL equation~\eqref{2dTL}, 
namely
\begin{eqnarray}
\Delta^+_l\Delta^-_mQ_{l,m,n}
=V_{l,m-1,n+1}-V_{l+1,m-1,n}-V_{l,m,n}+V_{l+1,m,n-1}\,,
\label{e:discrete2dToda}\\
V_{l,m,n}=(\delta \kappa)^{-1}\log [1+\delta \kappa\, (\exp Q_{l,m,n}-1)]\,,
\nonumber
\end{eqnarray}
with $l,m,n \in \mathbb Z$, 
$l$ and $m$ being the discrete analogues of 
the time~$t$ and space $x$ coordinates, respectively,
and where $\Delta^+_l$ and $\Delta^-_m$ are the forward and backward
difference operators defined by
\begin{eqnarray}
\Delta^+_l f_{l,m,n}=
\frac{f_{l+1,m,n}-f_{l,m,n}}{\delta},
\label{diffop1}
\\ 
\Delta^-_m f_{l,m,n} =
\frac{f_{l,m,n}-f_{l,m-1,n}}{\kappa}.
\label{diffop2}
\end{eqnarray}
Equation~\eqref{e:discrete2dToda}, which is the discrete analogue of Eq.~\eqref{2dTL},
can be written in bilinear form~\cite{Hirota2d} in a manner similar to~Eq.~\eqref{bilinear2dTL}:
\begin{eqnarray}
 (\Delta^+_l\Delta^-_m\tau_{l,m,n})\,\tau_{l,m,n} -
   (\Delta^+_l\tau_{l,m,n})\,\Delta^-_m\tau_{l,m,n}
\nonumber\\\kern6em
 = \tau_{l,m-1,n+1}\tau_{l+1,m,n-1} -
  \tau_{l+1,m-1,n}\tau_{l,m,n},
\label{toda}
\end{eqnarray}
with $Q_{l,m,n}$ related to %the $\tau$~function
$\tau_{l,m,n}$ by the transformation
$V_{l,m,n}=\Delta^+_l\Delta^-_m\log\tau_{l,m,n}$, i.e., 
\begin{equation}
Q_{l,m,n}=\log \frac{\tau_{l+1,m+1,n-1}\tau_{l,m,n+1}}
  {\tau_{l+1,m,n}\tau_{l,m+1,n}}\,.
\end{equation}
Note that $Q_{l,m,n}=\log[1+(\e^{\delta\kappa\,V_{l,m,n}}-1)/\delta\kappa]$.
Special solutions of Eq.~\eqref{toda} 
(which is the discrete analogue of Eq.~\eqref{bilinear2dTL})
are obtained when
the $\tau$~function $\tau_{l,m,n}$ is expressed in terms of
a Casorati determinant $\tau_{l,m,n}=\tau_{l,m,n}^{(\N)}$
as~\cite{Hirota2d}
\begin{equation}
\tau_{l,m,n}^{(\N)} = \left|
\begin{array}{cccc}
f^{(1)}_{l,m,n} & f^{(1)}_{l,m,n+1} & \cdots & f^{(1)}_{l,m,n+\N-1}\\ 
f^{(2)}_{l,m,n} & f^{(2)}_{l,m,n+1} & \cdots & f^{(2)}_{l,m,n+\N-1}\\ 
\vdots & \vdots & & \vdots \\
f^{(\N)}_{l,m,n} & f^{(\N)}_{l,m,n+1} & \cdots & f^{(\N)}_{l,m,n+\N-1}
\end{array}\right|, 
\label{tau}
\end{equation}
where each of the functions $\{f^{(i)}_{l,m,n}, i=1,2,\cdots,\N\}$ 
satisfies the following discrete dispersion relations:
\begin{eqnarray}
\Delta^+_l f_{l,m,n}= f_{l,m,n+1},
\label{toda:disp}
\\
\Delta^-_m f_{l,m,n}= - f_{l,m,n-1}.
\label{toda:disp2}
\end{eqnarray}
If we take as a solution for Eqs.~\eqref{toda:disp} and~\eqref{toda:disp2}
the functions
\begin{eqnarray}
f^{(i)}_{l,m,n} = \phi(p_i) + \phi(q_i)\,,
\label{eq:rei1}
\\
\noalign{\noindent with}
\phi(p)= p^n (1+\delta p)^l (1+\kappa p^{-1})^{-m}\,,
\label{e:phidef}
\end{eqnarray}
the $\tau$~function~\eqref{tau} yields a $\N$-soliton solution
for the discrete 2DTL~Eq.~\eqref{toda}.

As in the semi-continuous 2DTL, however, solutions of Eq.~\eqref{toda}
can also be obtained when we
consider the $\tau$~function defined by the Hankel determinant
\begin{equation}
\tau_{l,m,n}^{(\N)} = \left|
\begin{array}{cccc}
f_{l,m,n} & f_{l,m,n+1} & \cdots & f_{l,m,n+\N-1}\\ 
f_{l,m,n+1} & f_{l,m,n+2} & \cdots & f_{l,m,n+\N}\\ 
\vdots & \vdots & & \vdots \\
f_{l,m,n+\N-1} & f_{l,m,n+\N} & \cdots & f_{l,m,n+2\N-2}
\end{array}\right|\,, 
\label{tau3}
\end{equation}
where
\begin{eqnarray}
\label{e:discrete2dTLfdef}
f_{l,m,n}= 
\sum_{i=1}^\M \alpha_i\, \phi(p_i)\,,
\\
\noalign{\noindent which corresponds to choosing}
f_{l,m,n}^{(i)}= f_{l,m,n+i-1}\,,
%p_i^n (1+\delta p_i)^l (1+\kappa p_i^{-1})^{-m},
\label{eq:rei13}
\end{eqnarray}
for $i=1,\dots,\N$. 
Without loss of generality, we can label the parameters $p_i$ so that
$0 < p_1 < p_2 < \cdots <p_{\M-1}< p_\M$. 
Then, as in the semi-continuous 2DTL, we have the following:
\begin{lemma}
\label{d-2dTL:tauN}
Let $f_{l,m,n}$ be given by Eq.~\eqref{eq:rei13}.  
Then, for $1\leq \N\leq \M-1$, 
the $\tau$~function defined by the Hankel determinant~\eqref{tau3} 
has the form 
\begin{equation}
\tau^{(\N)}_{n}=
  \kern-1em \sum_{1\leq i_1< \cdots < i_\N\leq \M}
  \Delta(i_1,\dots ,i_\N) 
  \prod_{j=1}^\N\alpha_{i_j} \, \phi(p_{i_j})
%    p_{i_j}^n (1+\delta p_{i_j})^l (1+\kappa p_{i_j}^{-1})^{-m}
\label{d-tauplus}
\end{equation}
where $\Delta(i_1,\dots ,i_\N)$ is the square of the 
van der Monde determinant,
\begin{equation}
\Delta(i_1,\dots ,i_\N)=\prod_{1\leq j<l\leq \N}(p_{i_j}-p_{i_l})^2,.
\end{equation}
\end{lemma}
\begin{proof}
Again, the result follows by applying the Binet-Cauchy theorem to the
Hankel determinant~\eqref{tau3}.
\end{proof}

Unlike its counterpart in the semi-continuous 2DTL equation, 
the $\tau$~function in Eq.~\eqref{tau3} cannot be written 
in terms of a Wronskian, since no derivatives appear.
However, as in the semi-continuous 2DTL equation,
the $\tau$~function thus defined is positive definite, and therefore 
all the solutions generated by it are non-singular.
In the next section we show that, like its analogue in the 
semi-continuous 2DTL equation,
the above $\tau$~function produces soliton solutions 
of resonant type with web structure,
and we conjecture that, like in the continous case,
an $(\Np,\Nm)$-soliton with $\Np=\M-\N$ and $\Nm=\N$ is created.

Like with the semi-continuous 2DTL equation, however, 
before discussing resonant solutions it is convenient to 
first look at one-soliton solutions of the fully discrete 2DTL equation.
Let us introduce the analogue of Eq.~\eqref{e:wndef}, namely the function
\begin{equation}
w_{l,m,n}= \log \frac{\tau_{l,m+1,n-1}}{\tau_{l,m,n}}\,,
\label{e:wlmndef}
\end{equation}
so that the solution of the discrete 2DTL equation is given by
\begin{equation}
Q_{l,m,n}=\log 
\frac{\tau_{l+1,m+1,n-1}\tau_{l,m,n+1}}
{\tau_{l+1,m,n}\tau_{l,m+1,n}}=
w_{l+1,m,n}-w_{l,m,n+1}\,.
\label{d:1-soliton}
\end{equation}
It is also useful to rewrite the function $\phi(p_i)$ in 
Eqs.~\eqref{e:phidef}, \eqref{e:discrete2dTLfdef}
as $\phi(p_i)=e^{\theta_{l,m,n}^{(i)}}$, where
\begin{equation}
\theta_{l,m,n}^{(i)}=n\log p_i-m\log (1+\kappa p_i^{-1})
+l\log (1+\delta p_i)+\theta^{(i)}_0\,,
\label{e:discretethetadef}
\end{equation}
If 
$\tau_n=e^{\theta_{l,m,n}^{(1)}}+e^{\theta_{l,m,n}^{(2)}}$,
with $p_1<p_2$, 
%where 
%$\theta_{l,m,n}^{(i)}=n\log p_i-m\log (1+\kappa p_i^{-1})
%+l\log (1+\delta p_i)+\theta^{(i)}_0$, 
then $w_{l,m,n}$ is given by
\[
%%%\begin{eqnarray*}
\fl
\begin{array}{lllll}
w_{l,m,n}&=&
\log \frac{1}{2}
(p_1^{-1}(1+\kappa p_1^{-1})^{-1}+p_2^{-1}(1+\kappa p_2^{-1})^{-1})\\
&\quad &\quad +\frac{1}{2}
(p_1^{-1}(1+\kappa p_1^{-1})^{-1}-p_2^{-1}(1+\kappa p_2^{-1})^{-1})
{\rm tanh}\frac{1}{2}(\theta_{l,m,n}^{(1)}-\theta_{l,m,n}^{(2)})\\
[1.2ex]
&\longrightarrow&
\left\{
\begin{array}{ll}
-\log p_1-\log (1+\kappa p_1^{-1}) &\quad{\rm as}~~n\to\infty,\\
-\log p_2-\log (1+\kappa p_2^{-1}) &\quad{\rm as}~~n\to -\infty,
\end{array}
\right. 
\end{array}
\]
which leads to the one-soliton solution of the discrete 2DTL equation.
In the $n$-$m$ plane, this solution describes a plane wave
$\exp(Q_{l,m,n})=\Phi({\bf k}\cdot{\bf x}-\omega\,l)$
with ${\bf x}=(n,m)$,
having wavenumber vector ${\bf k}=(k_n,k_m)$ and frequency $\omega$
given by:
\begin{eqnarray*}
&&{\bf k}=(\log p_1-\log p_2,~
-\log (1+\kappa p_1^{-1})+\log (1+\kappa p_2^{-1}))=:{\bf k}_{1,2},\quad
\\
&&\omega=-\log (1+\delta p_1)+\log (1+\delta p_2)=:\omega_{1,2}.
\end{eqnarray*}
The soliton parameters $({\bf k},~\omega)$ now satisfy the discrete
dispersion relation
$(e^{-\omega}-1)(1-e^{-k_m})=\delta \kappa (e^{-k_m+k_n}
+e^{-\omega-k_n}-e^{-\omega-k_m}-1)$. 
The one-soliton solution~\eqref{d:1-soliton} is referred to as a 
line soliton since, like its semi-continuous analogue, it is localized 
around the (contour) line $\theta_{l,m,n}^{(1)}=\theta_{l,m,n}^{(2)}$ 
in the $n$-$m$ plane.
Again, we will refer to $c=dn/dm$ as the velocity of the line soliton 
in the $n$ direction.
For the above line soliton solution, this velocity is given by $c_{1,2}$,
where now
\begin{equation}
c_{i,j}=(\log (1+\kappa p_i^{-1})-\log (1+\kappa p_j^{-1}))
/(\log p_i-\log p_j)\,.
\label{e:discretesolitonvelocity}
\end{equation}

%%%%%%%%%%%%%%%%%%%%%%%%%%%%%%%%%%%%%%%%%%%%%%%%%%%%%%%%%%%%%%%%%%%%%%%%%
\section{Resonance and web structure in the discrete 2D Toda lattice equation}

As in the semi-continuous case, we first consider $(\Nm,1)$-soliton solutions,
i.e., solutions obtained in the case $\N=1$,
and in particular we start from (2,1)-soliton solutions
(i.e., the case $\N=1$ and $\M=3$), 
whose $\tau$-function is given by
\[
\tau_n= e^{\theta_{l,m,n}^{(1)}} 
+ e^{\theta_{l,m,n}^{(2)}} + e^{\theta_{l,m,n}^{(3)}}\,,
\]
with $\theta_{l,m,n}^{(i)}$ given by Eq.~\eqref{e:discretethetadef},
and where $p_1<p_2<p_3$ without loss of generality.
%$\theta_{l,m,n}^{(i)}=n\log p_i-m\log (1+\kappa p_i^{-1})
%+l\log (1+\delta p_i)+\theta^{(i)}_0$. 
As in the continous case, 
this solution describes the confluence of two shocks:
two shocks for $m\to \infty$ 
(each corresponding to a line soliton for $Q_{l,m,n}$)
with velocities $c_{1,2}$ and $c_{1,3}$
merge into a single shock for $m\to -\infty$ with velocity $c_{1,3}$,
where $c_{i,j}$ is given by Eq.~\eqref{e:discretesolitonvelocity} in all cases.
%$c_{i,j}=(\log (1+\kappa p_i^{-1})-\log (1+\kappa p_j^{-1}))
%/(\log p_i-\log p_j)$.
%$c_{1,2}=(\log (1+\kappa p_1^{-1})-\log (1+\kappa p_2^{-1}))
%/(\log p_1-\log p_2)$ and
%$c_{2,3}=(\log (1+\kappa p_2^{-1})-\log (1+\kappa p_3^{-1}))
%/(\log p_2-\log p_3)$, 
%and the single shock for $m\to\infty$
%has velocity $c_{1,3}=(\log (1+\kappa p_1^{-1})-\log (1+\kappa p_3^{-1}))
%/(\log p_1-\log p_3)$.
%
This Y-shape interaction represents a resonance of three line solitons.
The resonance conditions for three solitons with the wavenumber vectors
$\{{\bf k}_{i,j}\,|~1\le i<j\le 3\}$ and the frequencies
$\{\omega_{i,j}\,|~1\le i<j\le 3\}$
are still given by Eq.~\eqref{resonancecondition},
and again are trivially satisfied by those line solitons.
Furthermore, this solution is also the resonant case
of the ordinary 2-soliton solution of the discrete 2DTL equation,
arising in the limit of an infinite phase shift.
The resonance process for the $(\Np,1)$-soliton solutions of the 
discrete 2DTL equation can be expressed as a generalization
of the confluence of shocks discussed earlier. 

Next we consider more general $(\Np,\Nm)$-soliton solutions.  
Following Ref.~\cite{JPhysA2003v36p10519} and the semi-continuous case, 
we can describe the asymptotic pattern of the solution
by introducing a local coordinate frame $(\xi,m)$ in order
to study the asymptotics for large $|m|$ with
\[
n= c\,m + \xi\,.
\]
Then the phase functions $\theta_{l,m,n}^{(i)}$ 
%in $f_{l,m,n}=\sum_{j=1}^{\M}e^{\theta_{l,m,n}^{(j)}}$
%($=\sum_{j=1}^{\M}\alpha_i p_i^n (1+\delta p_i)^l
%(1+\kappa p_i^{-1})^{-m}$) 
become
\[
\theta_{l,m,n}^{(i)}
=\xi \log p_i + \eta_i(c)\,m + \theta_0^{(i)}\,, \quad{\rm for}\quad
i=1,\dots, \M\,,
\]
with
\[
\eta_i(c):=\log p_i\, (c-\log(1+\kappa p_i^{-1})/\log p_i)\,.
\]
Without loss of generality, we assume an ordering for the
parameters $\{p_i\,|~i=1,\dots,\M\}$:
$0<p_1<p_2<\cdots<p_\M\,$.
Then, as in the semi-continuous case, 
one can easily show that the lines $\eta=\eta_i(c)$ are in general
position. 
%that is, each line $\eta=\eta_i(c)$ intersects with all other
%lines at $\M-1$ distinct points in the $c$-$\eta$ plane;
%in other words, only {\it two} lines meets at each intersection point.
As before, the goal is then to find the dominant exponential terms in the
$\tau_{l,m,n}^{(\N)}$-function~\eqref{d-tauplus} 
for $m\to \pm\infty$ as a function of
the velocity~$c$. 
First note that if only one exponential is dominant, then
$w_{l,m,n}=\log (\tau_{l,m+1,n-1}^{(\N)}/\tau_{l,m,n}^{(\N)})$ 
is just a constant,
and therefore the solution $Q_{l,m,n}=w_{l+1,m,n}-w_{l,m,n+1}$ is zero.
Then, as in the semi-continuous case, nontrivial contributions to $Q_{l,m,n}$
arise when one can find
{\it two} exponential terms which dominate over the others.
Also, as in the semi-continuous case, 
since the intersections of the $\eta_i$'s are always pairwise,
three or more terms cannot make a dominant balance for large $|m|$.
For $(\Nm,1)$-soliton solutions, it is easy to see that
at each $c$ the dominant exponential term for $m\to -\infty$ is provided
by only $\eta_1$ and/or $\eta_\M$, and therefore there is only one shock
($\N=1$) moving with velocity~
$c_{1,\M}$
%=(\log (1+\kappa p_{\M}^{-1})-\log (1+\kappa p_{1}^{-1}))
%/(\log p_{\M}-\log p_1)$ 
corresponding to
the intersection point of $\eta_1$ and $\eta_\M$.
On the other hand, as $m\to \infty$, each term $\eta_j$ can become
dominant for some $c$,
and at each intersection point $\eta_j=\eta_{j+1}$
the two exponential terms corresponding to $\eta_j$ and $\eta_{j+1}$
give a dominant balance;
therefore there are $\Np=\M-1$ shocks moving with
velocities~$c_{j,j+1}$ 
%= (\log (1+\kappa p_{j+1}^{-1})-\log (1+\kappa p_{j}^{-1}))
%/(\log p_{j+1}-\log p_j)$, 
for $j=1,\dots,\M-1$.

In the general case, $\N\ne1$, the $\tau_{l,m,n}^{(\N)}$-function
in~\eqref{d-tauplus}
involves exponential terms having combinations of phases, and
two exponential terms that make a dominant balance
can be found in a similar way as in the semi-continuous case.
We define again the {\it level of intersection} of the $\eta_i(c)$.
Again, $c_{i,j}$ in Eq.~\eqref{e:discretesolitonvelocity} identifies
the intersection point of $\eta_i(c)$ and $\eta_j(c)$,
i.e., $\eta_i(c_{i,j})=\eta_j(c_{i,j})$.

\begin{definition}
We define the level of intersection, denoted by $\sigma_{i,j}$, 
as the number of other $\eta_l$'s that at $c=c_{i,j}$ are smaller
than $\eta_i(c_{i,j})=\eta_j(c_{i,j})$.
That is,
\[
\sigma_{i,j}:=\big|\{\eta_l\,|~\eta_l(c_{i,j})<\eta_i(c_{i,j})
=\eta_j(c_{i,j})\}\big|\,.
\]
We also define $I(s)$ as the set of pairs $(\eta_i,\eta_j)$ having
the level $\sigma_{i,j}=s$, namely
\[ 
I(s):=\{(\eta_i,\eta_j)\,|~\sigma_{i,j}=s,~{\rm for}~i<j\,\}.
\]
\end{definition}

\noindent
As in the semi-continuous case, one can then show the following:

\begin{lemma}
\label{l:intersectingpairs-d}
The set $I(s)$ is given by
\[
I(s)=\{(\eta_i,\eta_{\M-s+i-1})\,|~i=1,\dots,s+1\}\,.
\]
\end{lemma}
\begin{proof}
See the proof of Lemma~\ref{l:intersectingpairs}.
\end{proof}

\iffalse
\begin{proof}
 From the assumption $q_1<q_2<\cdots<q_\M$, we have the following inequality
at $c=c_{i,j}$~ (i.e. $\eta_i=\eta_j$) for $i<j$,
\[
\eta_{i+1},\dots,\eta_{j-1}
<\eta_i=\eta_{j}
<\eta_1,\dots,\eta_{i-1},\,\eta_{j+1},\dots,\eta_\M\,.
\]
Then taking $j=\M-s-1$ leads to the assertion of the Lemma.
\end{proof} 
\fi

\noindent
As in the semi-continuous case, let $\Nm=\N$ and $\Np=\M-\N$.
The above lemma indicates that, for each intersecting pair $(\eta_i,\eta_j)$
with the level $\Np-1$ ($\N-1$),
there are $\Nm-1$ terms $\eta_l$'s which are larger (smaller)
than~$\eta_i=\eta_j$.
Then the sum of those $\Nm-1$ terms with either $\eta_i$ or $\eta_j$
provides two dominant exponents in the $\tau_{l,m,n}^{(\N)}$-function
for $m\to \infty$ $(m\to -\infty)$. 
We then have the following:
%(see more detail in the proof of
%Theorem~\ref{t:asymptotic-d}).
%Note also that $|I(N_\pm-1)|=N_\mp$.
%Now we can state our main theorem:

%%%%%%%%%%%%%%%%%%%%%%%%%%%%%%%%%%%%%%%%%%%%%%%%%%%%%%%%%%%%%%%%%%%%%%%%%
\begin{figure}[t!]
\centerline{
\includegraphics[scale=0.675]{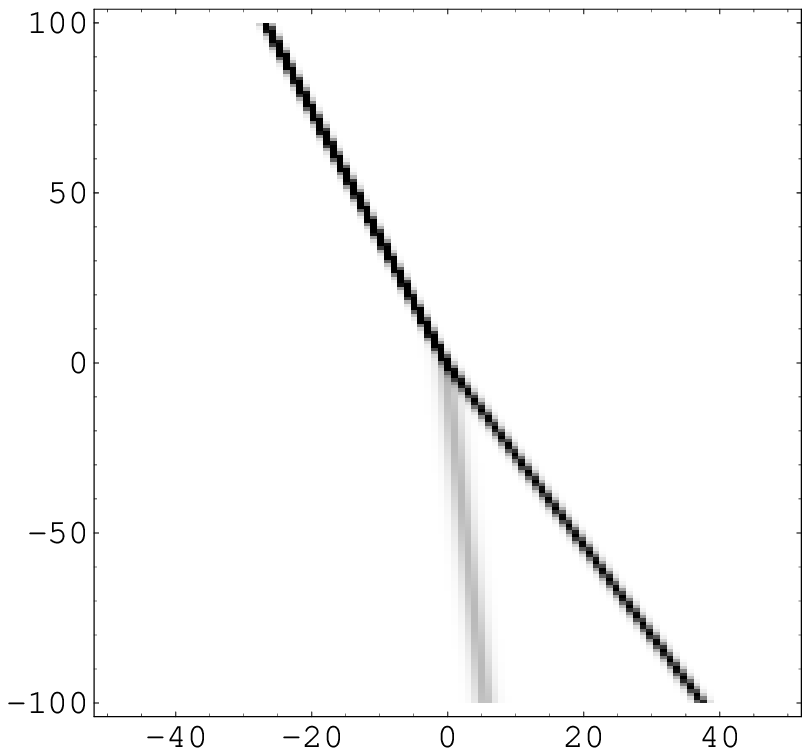}\quad
\includegraphics[scale=0.675]{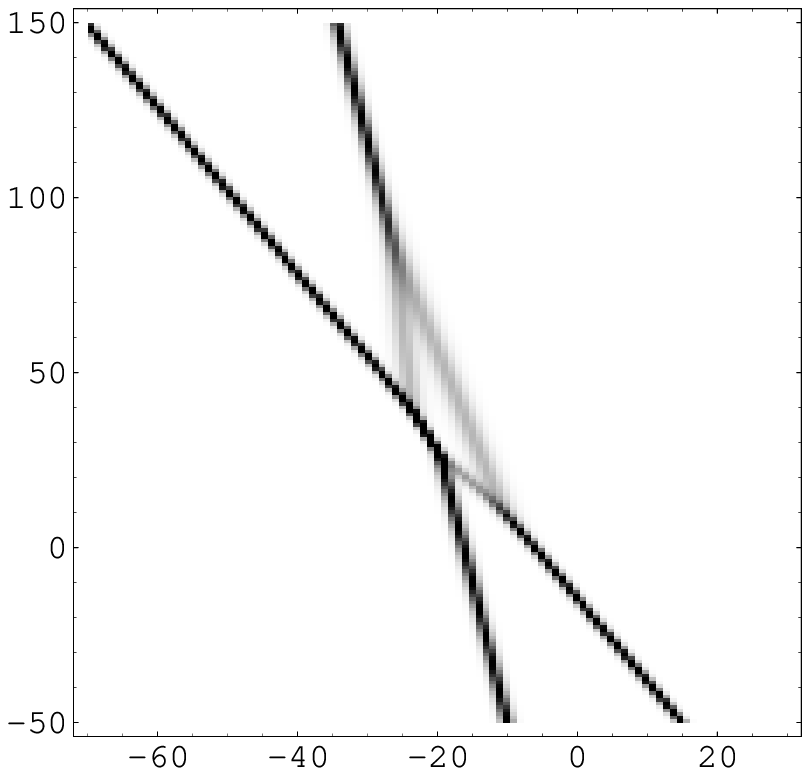}}
\kern-0.385\textwidth
\hbox to \textwidth{\hss(a)\kern0em\hss(b)\kern4em}
\kern+0.355\textwidth
\centerline{
\includegraphics[scale=0.675]{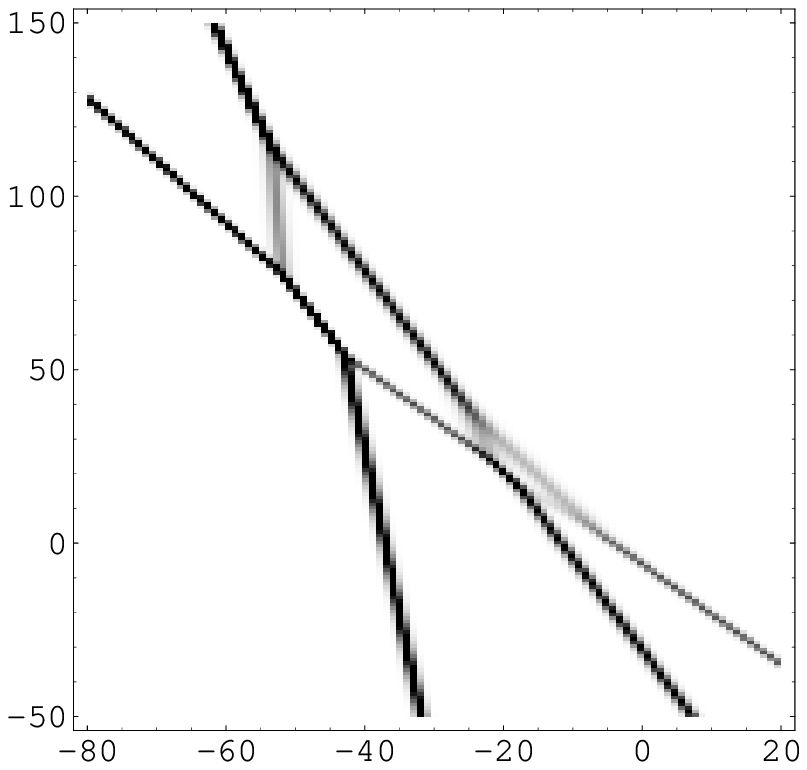}\quad
\includegraphics[scale=0.675]{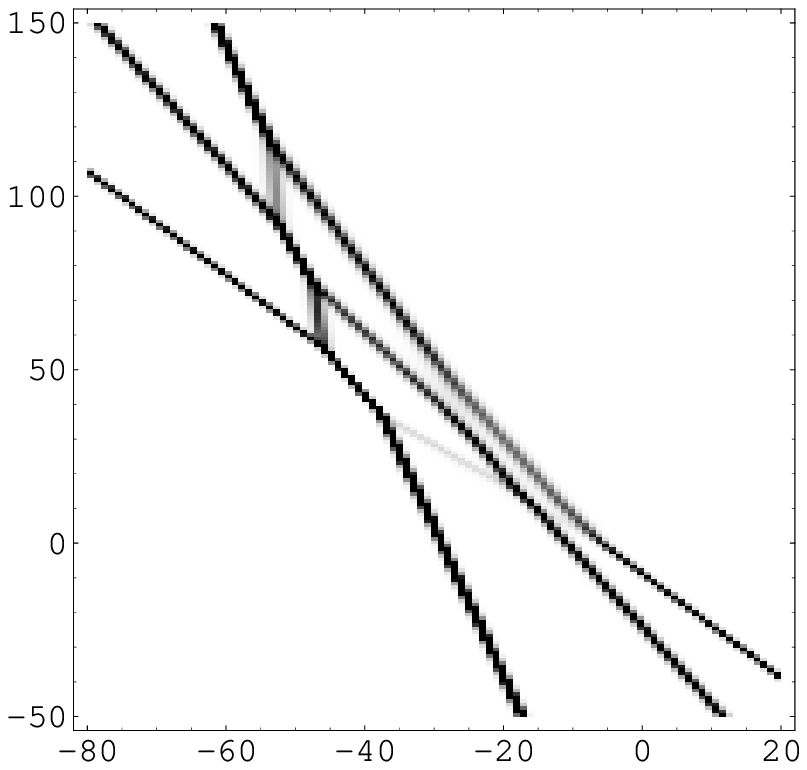}}
\kern-0.385\textwidth
\hbox to \textwidth{\hss(c)\kern0em\hss(d)\kern4em}
\kern+0.355\textwidth
\caption{Resonant solutions of the fully discrete two-dimensional 
Toda lattice:
(a)~(2,1)-soliton solution (i.e., a Y-junction) at $l=0$, 
with $p_1=1/10$, $p_2=1/2$, $p_3=10$, 
(b)~(2,2)-soliton solution at $l=40$, 
with $p_1=1/10$, $p_2=1/2$, $p_3=2$, $p_4=15$;
(c)~(3,2)-soliton solution at $l=80$, 
with $p_1=1/20$, $p_2=1/2$, $p_3=2$, $p_4=10$, $p_5=60$;
(d)~(3,3)-soliton solution at $l=80$,
with $p_1=1/20$, $p_2=1/2$, $p_3=2$, $p_4=10$, $p_5=20$, $p_6=120$.
In all cases $\delta=\kappa=1/4$;
the horizontal axis is $n$ and the vertical axis is $n$,
and each figure is a plot of $Q_{l,m,n}$ in logarithmic grayscale.
Note that the values of both $m$ and $n$ in the horizontal and 
vertical axes are discrete.}
\label{f:discrete2dtoda}
\end{figure}
%%%%%%%%%%%%%%%%%%%%%%%%%%%%%%%%%%%%%%%%%%%%%%%%%%%%%%%%%%%%%%%%%%%%%%%%%

\begin{theorem}
\label{t:asymptotic-d}
Let\, $w_{l,m,n}$\, be a function defined by Eq.~\eqref{e:wlmndef},
with $\tau_{l,m,n}^{(\N)}$ given by Eq.~\eqref{d-tauplus}.
Then\, $w_{l,m,n}$\, has the following asymptotics for $m\to\pm\infty:$
\renewcommand\labelenumi{{\rm(\roman{enumi})}}
\begin{enumerate}
\itemsep 0pt
\parsep 0pt
\item
For\, $m\to \infty$\, and\, $n=c_{i,\Nm+i}\,m+\xi$ ~for $i=1,\dots,\Np$\,,
\[
\kern-\leftmargini
w_{l,m,n}~\longrightarrow~\left\{\!\!
  \begin{array}{ll}
    K_i(+,+):= \sum_{j=i+1}^{\N+i} \log p_j 
      &\mathrm{as}~~ \xi\to\infty\,,\\
    K_i(-,+):= \sum_{j=i}^{\N+i-1} \log p_j 
      &\mathrm{as}~~ \xi\to-\infty\,.
  \end{array}\right.
\]
\item
For\, $m\to -\infty$\, and\, $n=c_{i,\Np+i}\,m+\xi$~ for $i=1,\dots,\Nm$\,,
\[
\fl
\kern-\leftmargini
w_{l,m,n}~\longrightarrow~\left\{\!\!
  \begin{array}{ll}
    K_i(+,-):=  \sum_{j=1}^{i-1} \log p_j
             + \sum_{j=1}^{\N-i+1} \log p_{\M-j+1} &\mathrm{as}~~
	       \xi\to \infty\,,\\
    K_i(-,-):= \sum_{j=1}^i \log p_j
             + \sum_{j=1}^{\N-1} \log p_{\M-j+i}   &\mathrm{as}~~
               \xi\to -\infty\,.
  \end{array}\right.
\]
\end{enumerate}
where\, $c_{i,j}$ is given by Eq.~\eqref{e:discretesolitonvelocity}.
%$c_{i,j}=(\log (1+\kappa p_{j}^{-1})-\log (1+\kappa p_{i}^{-1}))
%/(\log p_{j}-\log p_i)$.
\end{theorem}
\eject
\begin{proof}
Once the obvious modifications are made, 
the proof proceeds exactly like in the semi-continuous case,
namely Theorem~\ref{t:asymptotic}.
\end{proof}

Like its counterpart in the semi-continuous case, 
Theorem~\ref{t:asymptotic-d} determines the complete 
structure of asymptotic patterns of the solutions $Q_{l,m,n}$ 
given by~\eqref{tau3}. 
Indeed, Theorem~\ref{t:asymptotic-d} can be summarized as follows:
As $m\to \infty$, the function $w_{l,m,n}$ 
has $\Np$ jumps, moving with velocities
$c_{j,\Nm+j}$ for $j=1,\dots,\Np$;
as $m\to -\infty$, $w_{l,m,n}$ has $\Nm$ jumps, moving with velocities
$c_{i,\Np+i}$ for $i=1,\dots,\Nm$.
Since each jump represents a line soliton of $Q_{l,m,n}$,
the whole solution therefore represents an $(\Np,\Nm)$-soliton.
The velocity of each of the asymptotic line solitons in the 
$(\Np,\Nm)$-soliton
is determined from the $c$-$\eta$ graph of the levels of intersections.
Note that, given a set of $\M$ phases (as determined by the parameters
$p_i$ for $i=1,\dots,\M$), the same graph can be used for
any $(\Np,\Nm)$-soliton with $\Nm+\Np=\M$.
As an example, in Fig.~\ref{f:discrete2dtoda} we show 
a (2,1)-soliton solution, a (2,2)-soliton solution,
a (2,3)-soliton solution and a (3,3)-soliton solution. 

In particular, if $\M=2\N$, we have $\Np=\Nm=\N$, and 
Theorem~\ref{t:asymptotic-d}
implies that the velocities of the $\N$ incoming 
solitons are equal to those
of the $\N$ outgoing solitons. 
%%%(Here, we call solitons at $m\to \infty$
%%%($m\to -\infty$) incoming(outgoing) solitons.)
In the case of the ordinary multi-soliton solutions of 
the discrete 2DTL equation, the $\tau$-function~\eqref{tau} 
does not contain all the possible combinations of phases,
and therefore Theorem~\ref{t:asymptotic-d} should be modified. 
However, as in the semi-continuous case, the idea of using the 
levels of intersection is still applicable, and one can find 
that the asymptotic velocities for the ordinary $\N$-solitons 
generated by the Casorati determinant (\ref{tau}) 
with $f_{i}=e^{\theta_n^{(2i-1)}}+e^{\theta_n^{(2i)}}$ 
$i=1,\dots,\N$ and $p_1<p_2<\cdots<p_{2\N}$ are
$c_{2i-1,2i}=(\log (1+\kappa p_{2i}^{-1})-\log (1+\kappa p_{2i-1}^{-1}))
/(\log p_{2i}-\log p_{2i-1})$,
Note that, like in the semi-continuous case, 
these velocities are different from those of the resonant
$\N$-soliton solutions.

%%%As in the continuous case, however, a full characterization
%%%of the solution remains a problem for further study.
The resonant solutions of the fully discrete 2DTL
provide the basis for the construction of the resonant 
solution of the ultra-discrete 2DTL, as is shown in the
next two sections.

%%%%%%%%%%%%%%%%%%%%%%%%%%%%%%%%%%%%%%%%%%%%%%%%%%%%%%%%%%%%%%%%%%%%%%%%%
\section{The ultra-discrete two-dimensional Toda lattice}

We now turn our attention to an ultra-discrete analogue of 
the 2DTL equation.
Using Eqs.~\eqref{diffop1}~and~\eqref{diffop2},
we first write the 2DTL Eq.~\eqref{toda} in bilinear form as 
\begin{equation}\fl
(1-\delta \kappa)\,\tau_{l+1,m,n}\tau_{l,m+1,n}
- \tau_{l+1,m+1,n}\tau_{l,m,n} 
+ \delta \kappa\,\, \tau_{l,m,n+1}\tau_{l+1,m+1,n-1}
= 0\,,
\label{toda2}
\end{equation}
We define the difference operator $\Delta'$ as
\begin{equation}
\Delta' = \e^{-\partial_n} 
(\Delta^+_n-\Delta^+_l)(\Delta^+_n-\Delta^+_m)\,,
\label{e:deltaprime}
\end{equation}
where from here on the symbols $\Delta^+_l$, $\Delta^+_m$ and 
$\Delta^+_n$ will be used to denote the difference operators
\begin{equation}
 \Delta^+_l = \e^{\partial_l} - 1\,,\qquad
 \Delta^+_m = \e^{\partial_m} - 1\,,\qquad
 \Delta^+_n = \e^{\partial_n} - 1\,,
\label{e:newdifference}
\end{equation}
and where the shift operators $\e^{\partial_l}$, $\e^{\partial_m}$ and
$\e^{\partial_n}$ are defined by $\e^{\partial_n}f_{l,m,n}=f_{l,m,n+1}$ 
etc.\break
That is,
\begin{equation}
\Delta'f_{l,m,n}= f_{l+1,m+1,n-1}+f_{l,m,n+1}-f_{l+1,m,n}-f_{l,m+1,n}\,.
\end{equation}
Using Eqs.~\eqref{e:newdifference} and~\eqref{e:deltaprime}, 
we can rewrite Eq.~\eqref{toda2} as
\begin{equation}
 (1-\delta \kappa) + \delta \kappa \exp\left[\Delta'\log\tau_{l,m,n}\right]
 = \exp[\Delta^+_l\Delta^+_m\log\tau_{l,m,n}],
\label{Sijn}
\end{equation}
which becomes, taking a logarithm and applying $\Delta'$
(assuming $\delta\kappa\ne1$),
%on both sides of Eq.~\eqref{Sijn},
\begin{eqnarray}
%\fl
%\Delta' \log(1-\delta \kappa) + 
\Delta'\log\left[ 1 + \frac{\delta \kappa}{1-\delta \kappa }\exp
(\Delta' \log\tau_{l,m,n})\right] = \Delta^+_l 
\Delta^+_m \Delta' \log\tau_{l,m,n}\,.
\label{todaS}
\end{eqnarray}
We now take an ultra-discrete limit of Eq.~\eqref{todaS}
following Refs.~\eqref{2dTL}~\cite{Mori,Nagai}.
This is accomplished by choosing the lattice intervals as
\begin{equation}
\delta_\epsilon = \e^{-\r/\epsilon},\qquad
\kappa_\epsilon = \e^{-\s/\epsilon},
\end{equation}
where $\r,\s \in {\mathbb Z}_{\geq 0}$ are some predetermined integer constants,
and by defining
\begin{equation}
v^\epsilon_{l,m,n}= \Delta' \epsilon\log\tau_{l,m,n}^\epsilon\,,
\label{e:vepsilonlmn}
\end{equation}
Taking the limit $\epsilon \to 0^+$ in Eq.~\eqref{todaS}
and noting that 
$\lim_{\epsilon\to 0^+}\epsilon \log(1+\e^{X/\epsilon})=\max(0,X)$,
we then obtain
\begin{eqnarray}
\Delta^+_l \Delta^+_m v_{l,m,n} = 
\Delta'%\e^{-\partial_n} (\Delta^+_n - \Delta^+_t)(\Delta^+_n - \Delta^+_m)
\max(0,v_{l,m,n}-\r-\s), 
\label{rutoda}
\end{eqnarray}
where 
$v_{l,m,n}=\displaystyle{\lim_{\epsilon \to 0^+}}v^\epsilon_{l,m,n}$.
That is, using Eq.~\eqref{e:vepsilonlmn},
\begin{equation}
v_{l,m,n}= \Delta' \lim_{\epsilon\to0^+}\epsilon\, \tau^\epsilon_{l,m,n}\,.
\label{e:vlmn}
\end{equation}
Equation~\eqref{rutoda} is the ultra-discrete analogue of the 2DTL equation,
and can be considered a cellular automaton in the sense that 
$v_{l,m,n}$ takes on integer values.

Let us briefly discuss ordinary soliton solutions of the ultra-discrete
2DTL~Eq.~\eqref{rutoda}. 
As shown in Refs.~\cite{Mori,Nagai}, soliton solutions for 
the ultra-discrete 2DTL~Eq.~\eqref{rutoda} 
are obtained by an ultra-discretization of the soliton solution of
the discrete 2DTL~Eq.~\eqref{toda}.
For example, a one-soliton solution for Eq.~\eqref{toda} is given by
\begin{eqnarray}
\tau_{l,m,n}=1 + \eta_1,
\\ 
\noalign{\noindent with}
\eta_i =\alpha_i \frac{\phi(p_i)}{\phi(q_i)}\,,
\\
\noalign{\noindent and where}
\phi(p)= p^n (1+\delta p)^l (1+\kappa p^{-1})^{-m}
%\left(\frac{p_1}{q_1}\right)^n
%\left(\frac{1+\delta p_1}{1+\delta q_1}\right)^l
%\left(\frac{1+\kappa q_1^{-1}}{1+\kappa p_1^{-1}}\right)^{m}.
\end{eqnarray}
as before.
%in Eq.~\eqref{e:phidef}.
We introduce a new dependent variable
\begin{eqnarray}
\rho_{l,m,n}^\epsilon = \epsilon \log \tau_{l,m,n},
\label{e:rholmn}
\\
\noalign{\noindent and new parameters $P_1,Q_1,A_1\in\mathbb Z$ as}
\e^{P_1/\epsilon} = p_1, \qquad 
\e^{Q_1/\epsilon} = q_1,\qquad 
\e^{A_1/\epsilon} = \alpha_1.
\end{eqnarray}
%with $P_1,Q_1,A_1\in\mathbb Z$. 
Taking the limit $\epsilon \to 0^+$, we then obtain
\begin{eqnarray}
\rho_{l,m,n} = \max(0,\Theta_1)\,,
\label{e:rho1s}
\\
\noalign{\noindent where}
\Theta_i = A_i + n\,(P_i-Q_i) + 
   l\{\max(0,P_i-\r)-\max(0,Q_i-\r)\} 
\nonumber \\ \kern2em
   + m \{\max(0,-Q_i-\s)-\max(0,-P_i-\s)\}\,,
\label{e:Thetadef}
\end{eqnarray}
with 
$\e^{-\r/\epsilon} = \delta$ and 
$\e^{-\s/\epsilon} = \kappa$ as before,
and where 
$\rho_{l,m,n}=\lim_{\epsilon\to 0^+} \rho_{l,m,n}^\epsilon$.
According to Eqs.~\eqref{e:vlmn} and~\eqref{e:rholmn},
the one-soliton solution for Eq.~\eqref{rutoda} is then given by
\begin{eqnarray}
v_{l,m,n}= 
\rho_{l+1,m+1,n-1}+\rho_{l,m,n+1}-\rho_{l+1,m,n}-\rho_{l,m+1,n}\,.
\label{e:vfromrho}
\end{eqnarray}

Using a similar procedure we can construct a two-soliton solution.
Equation~\eqref{toda} admits a two-soliton solution given by
\begin{eqnarray}
&&\tau_{l,m,n} = 1 + \eta_1 + \eta_2 + \theta_{12}\eta_1\eta_2,
\\ 
\noalign{\noindent with}
&&\theta_{12} = \frac{(p_2-p_1)(q_1-q_2)}{(q_1-p_2)(q_2-p_1)},
%&&\eta_i = \alpha_i \left(\frac{p_i}{q_i}\right)^n
%\left(\frac{1+\delta p_i}{1+\delta q_i}\right)^l
%\left(\frac{1+\kappa q_i^{-1}}{1+\kappa p_i^{-1}}\right)^m\, 
\end{eqnarray}
and where $\eta_i= \alpha_i \phi(p_i)/\phi(q_i)$ as before
($i=1,2$).
In order to take the ultra-discrete limit of the above solution,
we suppose without loss of generality that the soliton parameters
satisfy the inequality
\begin{equation}
0 < p_1 < p_2 < q_2 < q_1\,.
\end{equation}
Introducing again the dependent variable
$\rho_{l,m,n}^\epsilon = \epsilon \log \tau_{l,m,n}$,
as well as integer parameters $P_i$, $Q_i$ and $A_i$ as
\begin{eqnarray}
\e^{P_i/\epsilon}=p_i,\qquad
\e^{Q_i/\epsilon}=q_i, \qquad
\e^{A_i/\epsilon} = \alpha_i
\end{eqnarray}
($i=1,2$), 
and taking the limit of small $\epsilon$, we obtain 
\begin{eqnarray}
\rho_{l,m,n}= \max(0,\Theta_1,\Theta_2,\Theta_1+\Theta_2+P_2-Q_2),
\end{eqnarray}
where $\Theta_i$ ($i=1,2$) was defined in Eq.~\eqref{e:Thetadef},
with $\rho_{l,m,n}=\lim_{\epsilon\to 0^+} \rho_{l,m,n}^\epsilon$
again,
and where $v_{l,m,n}$ is obtained from $\rho_{l,m,n}$ using 
Eq.~\eqref{e:vfromrho}.
Note that $P_1 < P_2 < Q_2 < Q_1$.

More in general, starting from Eqs.~\eqref{tau} and~\eqref{eq:rei1}
(with $0<p_1<p_2<\cdots<p_\M<q_\M<q_{N-1}<\cdots<q_1$)
and repeating the same construction, one obtains
the $N$-soliton solution of the ultra-discrete 2DTL Eq.~\eqref{rutoda} 
as~\cite{Nagai}
\begin{equation}
\rho_{l,m,n}= \max\limits_{\mu=0,1}\bigg[
  \sum_{1\le i\le \M}\mu_i\Theta_i 
    + \sum_{1\le i<i'\le \M}\mu_i\mu_{i'}(P_{i'}-Q_{i'})
  \bigg]
\end{equation}
where $\max_{\mu=0,1}$ indicates maximization over 
all possible combinations of the integers $\mu_i=0,1$, 
with $i=1,\dots,\M$.
Again, $v_{l,m,n}$ is obtained from $\rho_{l,m,n}$ via
Eq.~\eqref{e:vfromrho}.

Ordinary soliton solutions corresponding to the above choices were
presented in Ref.~\cite{Mori,Nagai}.
In the next section we show how this basic construction 
can be generalized to obtain resonant soliton solutions.

%%%%%%%%%%%%%%%%%%%%%%%%%%%%%%%%%%%%%%%%%%%%%%%%%%%%%%%%%%%%%%%%%%%%%%%%%
\section{Resonance and web structure in the ultra-discrete 2D Toda lattice equation}

Following Ref.~\cite{JPhysA2003v36p10519}, 
we now construct more general solutions of the ultra-discrete 
2DTL equation~\eqref{rutoda}
which display soliton resonance and web structure.

%%%%%%%%%%%%%%%%%%%%%%%%%%%%%%%%%%%%%%%%%%%%%%%%%%%%%%%%%%%%%%%%%%%%%%%%%
\begin{figure}[t!]
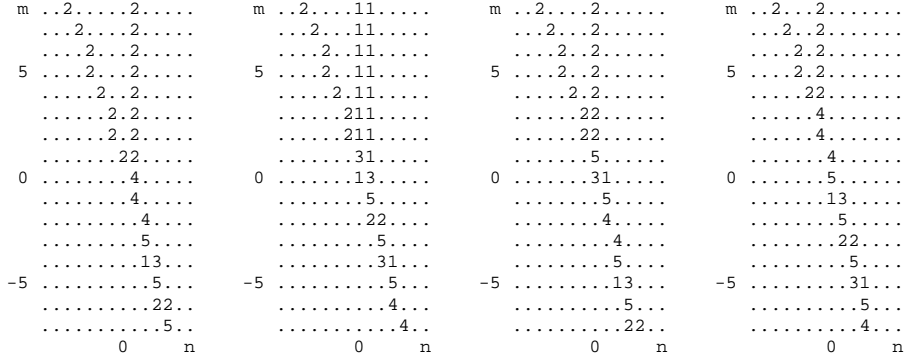

\begin{center}
\input fig3a
\input fig3b
\input fig3c
\input fig3d
\end{minipage}
\end{center}
\caption{A~$(2,1)$-resonant soliton solution (i.e., a Y-junction)
for the ultradiscrete 2DTL~Eq.~\eqref{rutoda}, with 
$P_1= -5$, $P_2=1$, $P_3 = 4$, $\r= 3$, $\s= 1$, $A_1-A_3=1$, $A_2-A_3=1$.
From left to right, the four plots represent the solution respectively at
$l=0$, $l=2$, $l=4$ and $l=6$.
The dots indicate zero values of $v_{l,m,n}$.}
\label{Y-soliton}
\end{figure}
%%%%%%%%%%%%%%%%%%%%%%%%%%%%%%%%%%%%%%%%%%%%%%%%%%%%%%%%%%%%%%%%%%%%%%%%%

%%%%%%%%%%%%%%%%%%%%%%%%%%%%%%%%%%%%%%%%%%%%%%%%%%%%%%%%%%%%%%%%%%%%%%%%%
\begin{figure}[t!]
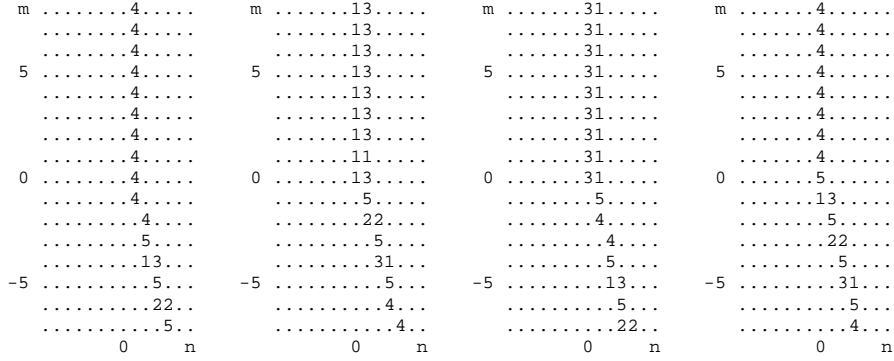

\begin{center}
\input fig4a
\input fig4b
\input fig4c
\input fig4d
\end{center}
\caption{An L-shape $(2,1)$-resonant soliton solution for Eq.~\eqref{rutoda}, with
$P_1 = -5$, $P_2=-1$, $P_3 = 4$, $ \r= 3$, $ \s= 1$, $A_1-A_3=1$, $A_2-A_3=1$.
From left to right, the four plots represent the solution respectively at
$l=0$, $l=2$, $l=4$ and $l=6$.}
\label{V-soliton}
\end{figure}
%%%%%%%%%%%%%%%%%%%%%%%%%%%%%%%%%%%%%%%%%%%%%%%%%%%%%%%%%%%%%%%%%%%%%%%%%

We first consider the case of a $(2,1)$-soliton for Eq.~\eqref{toda},
which is given by
\begin{equation}
\tau_{l,m,n} =\xi_1+\xi_2+\xi_3\,,
\end{equation}
where 
\begin{equation}
\xi_i = \alpha_i \phi(p_i)
%p_i^n (1+\delta p_i)^l (1+\kappa p_i^{-1})^{-m}
\end{equation}
($i=1,2,3$), with
\begin{equation}
\phi(p)= p^n (1+\delta p)^l (1+\kappa p^{-1})^{-m}
\label{e:phidef2}
\end{equation}
as before, and where again we take $0 < p_1 < p_2 < p_3$. 
As in the previous section,
we introduce the new dependent variable 
\begin{equation}
\rho_{l,m,n}^\epsilon = \epsilon \log \tau_{l,m,n}\,,
\end{equation}
and new parameters as
\begin{equation}
\e^{P_i/\epsilon} = p_i, \qquad
\e^{A_i/\epsilon} = \alpha_i
\end{equation}
($i=1,2,3$),
with $\e^{-\r/\epsilon} = \delta$ and 
$\e^{-\s/\epsilon} = \kappa$ as before. 
Taking the limit $\epsilon \to 0^+$, we then obtain
\begin{equation}
\rho_{l,m,n} = \max(R_1,R_2,R_3)
\end{equation}
where 
$\rho_{l,m,n}=\lim_{\epsilon\to 0^+} \rho_{l,m,n}^\epsilon$
as before, but where now
\begin{equation}
R_i = A_i + n P_i + l \max(0,P_i-\r) -m \max(0,-P_i-\s)
\label{e:Thetadef2}
\end{equation}
($i=1,2,3$).
Figure~\ref{Y-soliton} shows that this solution,
which again can be called a $(2,1)$-soliton,
is a Y-shape solution.
%%%%%Note however that the $(2,1)$-soliton in Fig.~\ref{Y-soliton} 
%%%%%looks like a $(1,2)$-soliton, in the sense that 
%%%%there are 2 solitons for 
%%%%%large positive $m$ and only one for large negative $m$.
%%%%%In general,
%%%%%a $(\Nm,\Np)$-soliton of the discrete 2DTL 
%%%%%equation~\eqref{e:discrete2dToda} 
%%%%%leads to a $(\Np,\Nm)$-soliton of Eq.~\eqref{rutoda} when taking the 
%%%%%ultra-discrete limit.
We note that, interestingly, an L-shape solution can be 
obtained instead of Y-shape solution for different solution parameters.
An example of such a L-shape soliton is shown in Fig.~\ref{V-soliton}. 
No analogue of this solution exists in the 2DTL and its
fully discrete version.

Next, we consider the case of a $(2,2)$-soliton for Eq.~\eqref{toda} 
following Ref.~\cite{JPhysA2003v36p10519}. 
Let us consider the following $\tau$~function
\begin{equation}
\tau_{l,m,n} = \left|
\begin{array}{cccc}
f_{l,m,n} & f_{l,m,n+1} \\ 
f_{l,m,n+1} & f_{l,m,n+2}
\end{array}\right|\,, \label{tau2}
\end{equation}
where
\begin{equation}
f_{l,m,n} = \xi_1 + \xi_2 + \xi_3 + \xi_4\,,
\end{equation}
where $\xi_i$ ($i=1,\dots,4$) is again defined as in Eq.~\eqref{e:phidef2},
%\begin{equation}
%&&f_{l,m,n} = \sum_{j=1}^4\xi_j\,,\\
%\noalign{\noindent with}
%\xi_j= \alpha_1 p_i^n (1+\delta p_i)^l (1+\kappa p_i^{-1})^{-m}
%\end{eqnarray}
and where $0 < p_1 < p_2 < p_3< p_4$ holds. 
We introduce again the new parameters $\e^{P_k/\epsilon} = p_k$
and $\e^{A_k/\epsilon} = \alpha_k$ ($k=1,\dots,4$)
and the new dependent variable
$\rho_{l,m,n}^\epsilon = \epsilon \log \tau_{l,m,n}$.
Taking the limit $\epsilon \to 0^+$, we then obtain
\begin{equation}
\rho_{l,m,n}= \max\limits_{1\le i<j\le 4}(K_{ij}+2P_j)\,,
%\max(K_{12}+2P_2,K_{13}+2P_3,
%K_{14}+2P_4,
%K_{23}+2P_3,K_{24}+2P_4,
%K_{34}+2P_4)
\end{equation}
where 
$\rho_{l,m,n}=\lim_{\epsilon\to 0^+} \rho_{l,m,n}^\epsilon$,
as before, and
\begin{equation}
K_{ij}=R_i+R_j\,,
%&&R_i = A_i+ n P_i + l \max(0,P_i-\r)-m \max(0,-P_i-\s),
%&&R_2 = A_2+ n P_2 + l \max(0,P_2-\r)-m \max(0,-P_2-\s),\\
%&&R_3 = A_3+ n P_3 + l \max(0,P_3-\r)-m \max(0,-P_3-\s),\\
%&&R_4 = A_4+ n P_4 + l \max(0,P_4-\r)-m \max(0,-P_4-\s).
\end{equation}
%($i=1,\dots,4$).
and with $R_j$ given by Eq.~\eqref{e:Thetadef2} as before.
Figure~\ref{f:ud2dtoda(2,2)} 
shows the temporal evolution of a $(2,2)$-soliton solution.  
Note the appearance of a hole in Fig.~\ref{f:ud2dtoda(2,2)}.

%%%%%%%%%%%%%%%%%%%%%%%%%%%%%%%%%%%%%%%%%%%%%%%%%%%%%%%%%%%%%%%%%%%%%%%%%
\begin{figure}[t!]
%\begin{center}
%\subfigure[$l=-12$]{\includegraphics[scale=0.5]{(2,2)-time-12.eps}}
%\subfigure[$l=-7$]{\includegraphics[scale=0.5]{(2,2)-time-7.eps}}
%\subfigure[$l=4$]{\includegraphics[scale=0.5]{(2,2)-time4.eps}}
%\subfigure[$l=7$]{\includegraphics[scale=0.5]{(2,2)-time7.eps}}
%\end{center}
\centerline{
\includegraphics[scale=0.5]{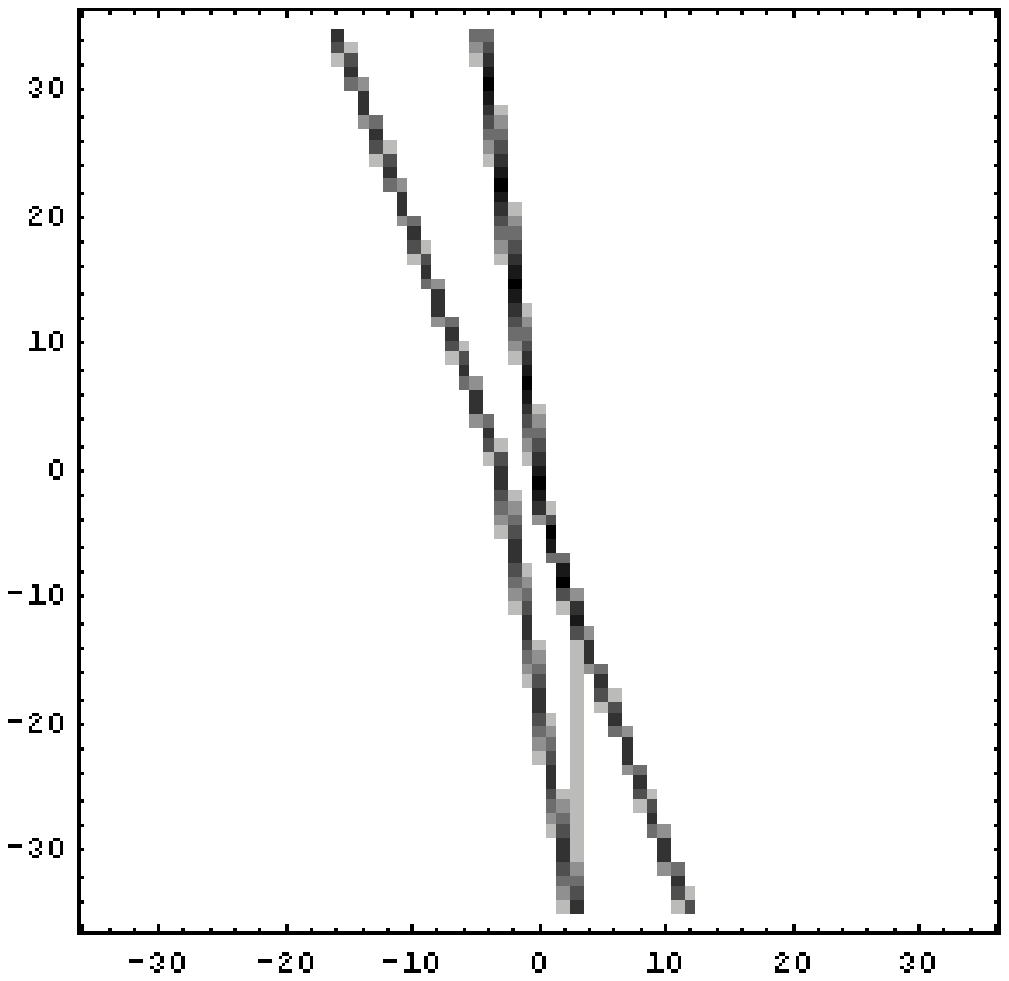}\quad 
\includegraphics[scale=0.5]{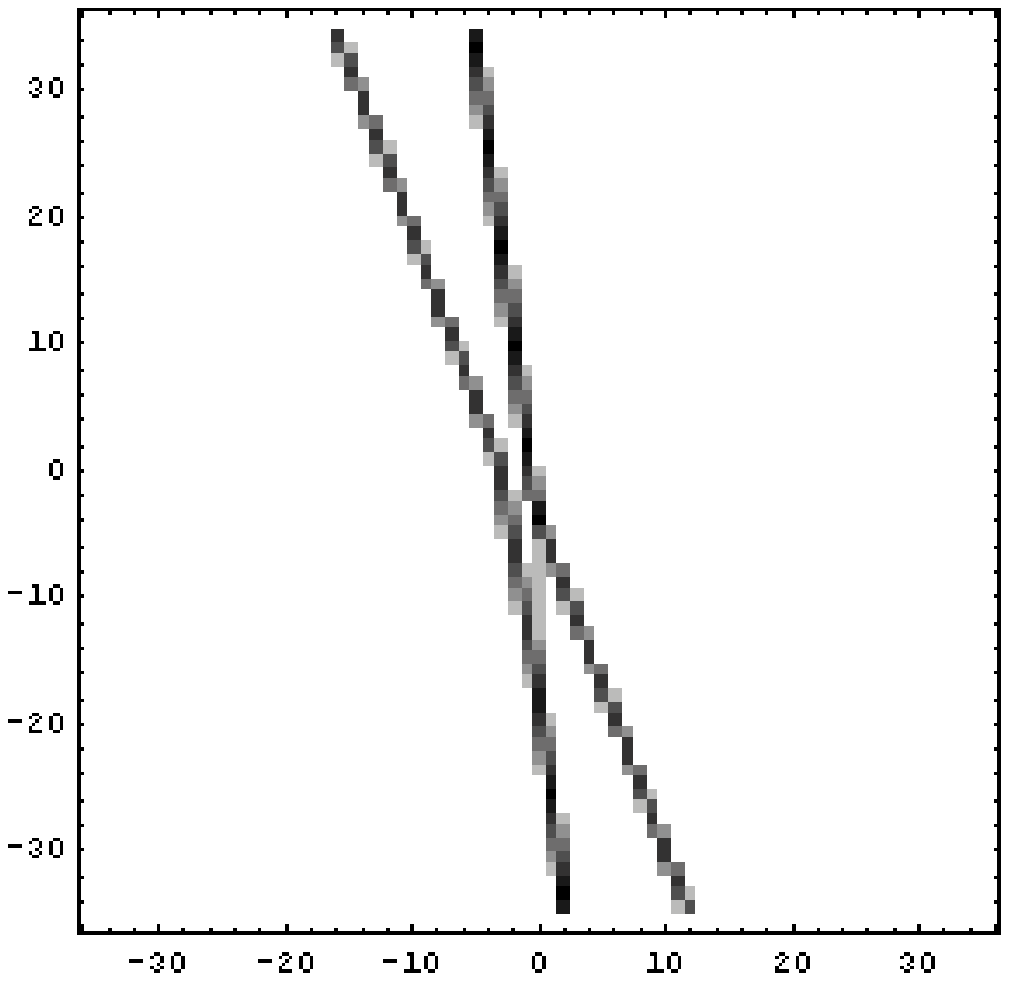}}
\kern-0.385\textwidth
\hbox to \textwidth{\hss(a)\kern0em\hss(b)\kern4em}
\kern+0.345\textwidth
\vglue\bigskipamount
\centerline{
\includegraphics[scale=0.5]{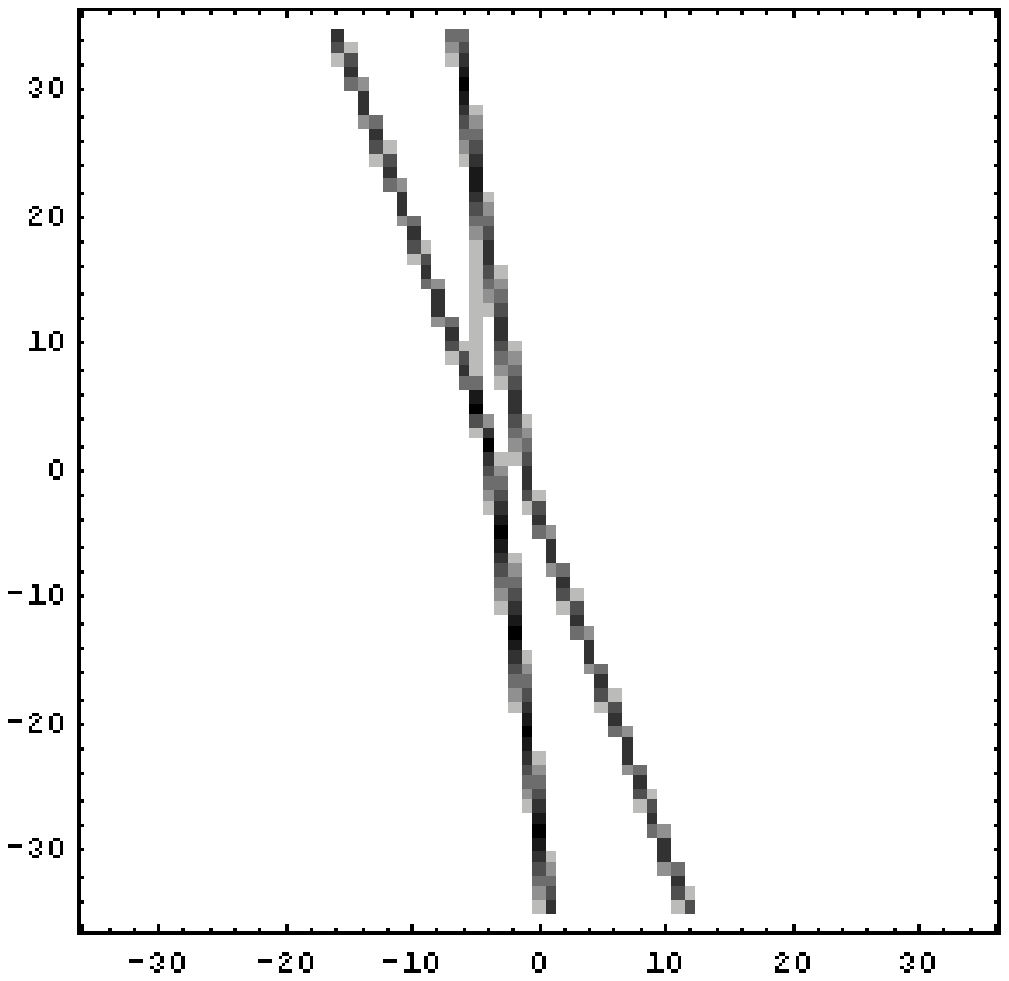}\quad\ 
\includegraphics[scale=0.5]{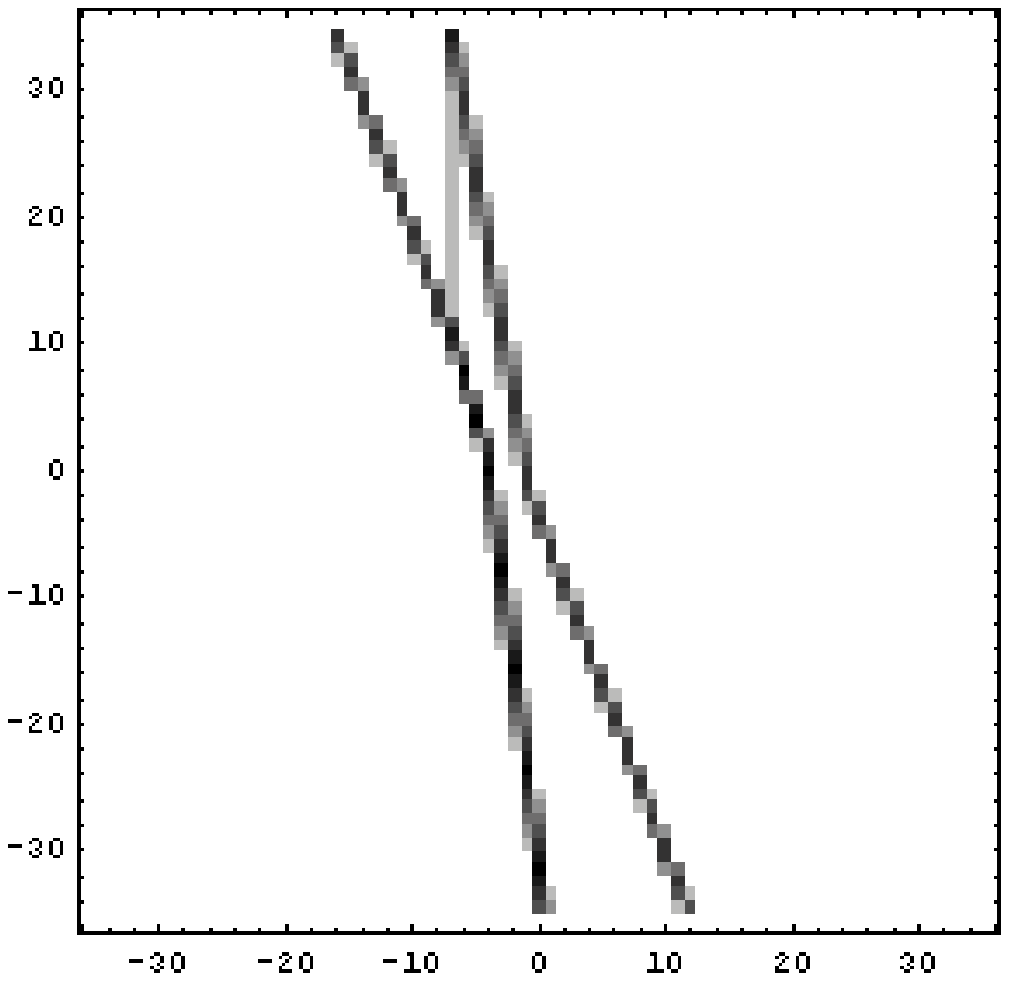}}
\kern-0.385\textwidth
\hbox to \textwidth{\hss(c)\kern0em\hss(d)\kern4em}
\kern+0.355\textwidth
\caption{Snapshots illustrating the temporal evolution of 
a $(2,2)$-resonant soliton solution for Eq.~\eqref{rutoda}, with
$P_1 = -7$, $P_2=-5$, $ P_3 = 1$, $P_4=3$, $ \r= 4$, $\s= 2$, 
$A_1=-10$, $A_2=-6$, $A_3=0$, $A_4=2$:
(a)~$l=-12$; (b)~$l=-7$; (c)~$l=4$; (d)~$l=7$.
As in Figs.~\ref{Y-soliton} and~\ref{V-soliton},
the horizontal axis is~$n$ and the vertical axis is~$m$.
Since the interaction extends over a wider range of values of $m$ and $n$,
the solution is now plotted in grayscale, in a similar way as in 
Figs.~\ref{f:2dtoda} and~\ref{f:discrete2dtoda}; 
the values of $v_{l,m,n}$ however are still discrete, 
as in Figs.~\ref{Y-soliton} and~\ref{V-soliton}.}
\label{f:ud2dtoda(2,2)}
\end{figure}
%%%%%%%%%%%%%%%%%%%%%%%%%%%%%%%%%%%%%%%%%%%%%%%%%%%%%%%%%%%%%%%%%%%%%%%%%

%%%%%%%%%%%%%%%%%%%%%%%%%%%%%%%%%%%%%%%%%%%%%%%%%%%%%%%%%%%%%%%%%%%%%%%%%
\begin{figure}[t!]
%\begin{center}
%\subfigure[$l=-10$]{\includegraphics[scale=0.5]{time-10.eps}}
%\subfigure[$l=0$]{\includegraphics[scale=0.5]{time0.eps}}
%\subfigure[$l=10$]{\includegraphics[scale=0.5]{time10.eps}}
%\end{center}
\centerline{
\includegraphics[scale=0.5]{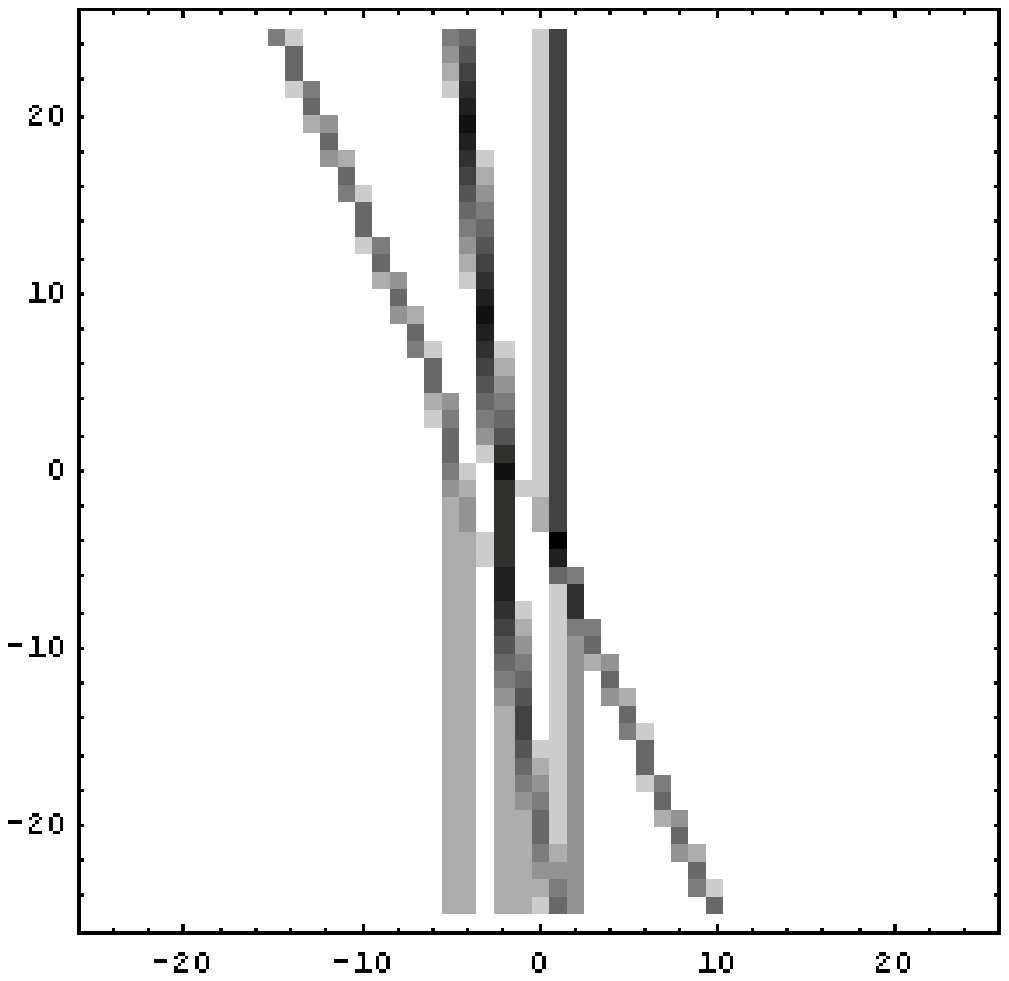}\quad\ 
\includegraphics[scale=0.5]{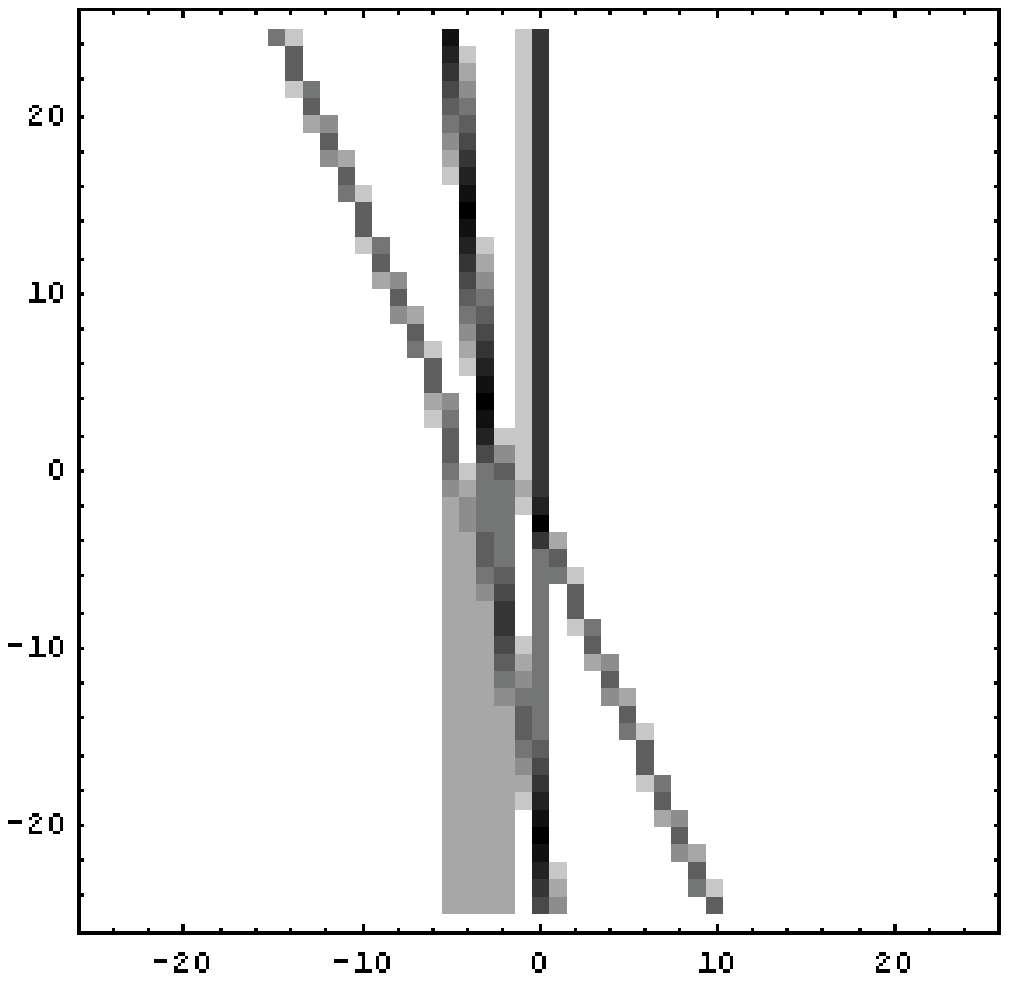}}
\kern-0.385\textwidth
\hbox to \textwidth{\hss(a)\kern0em\hss(b)\kern4em}
\kern+0.345\textwidth
\vglue\bigskipamount
\centerline{
\includegraphics[scale=0.5]{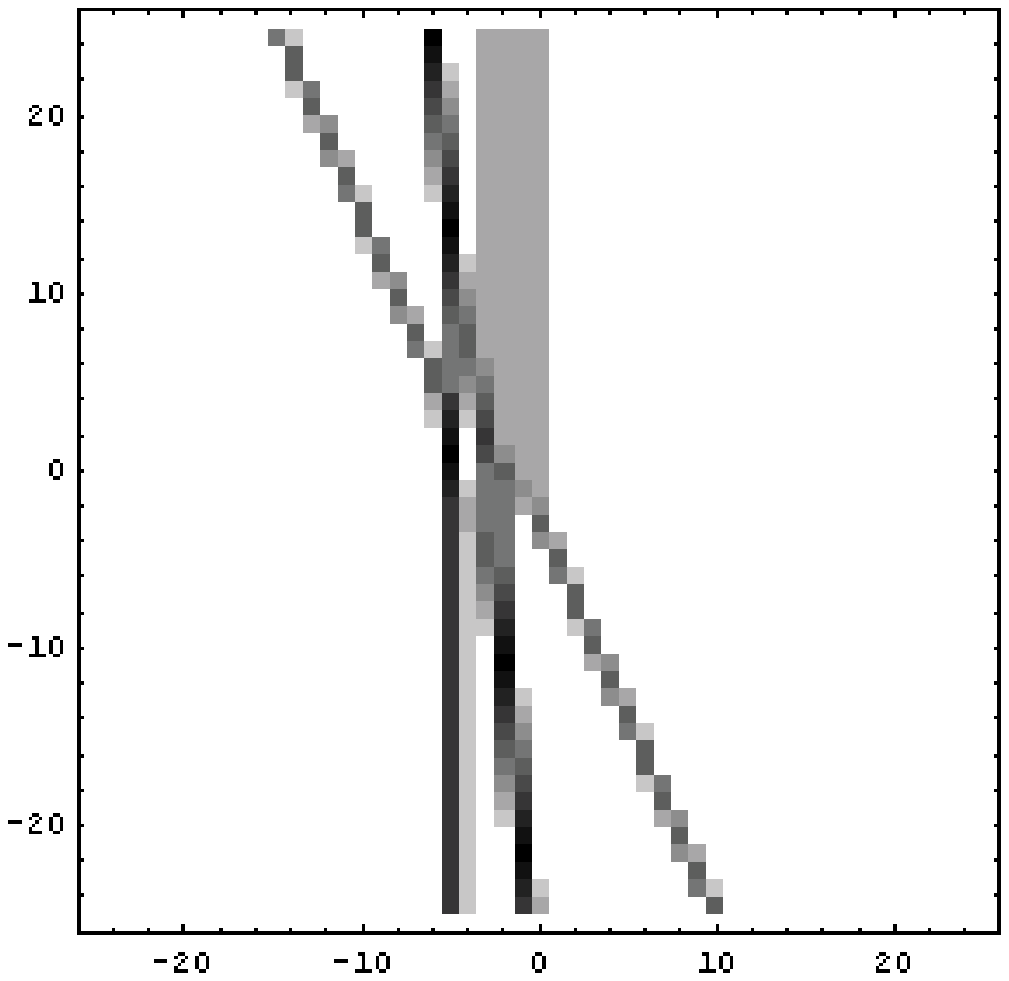}\quad\ 
\includegraphics[scale=0.5]{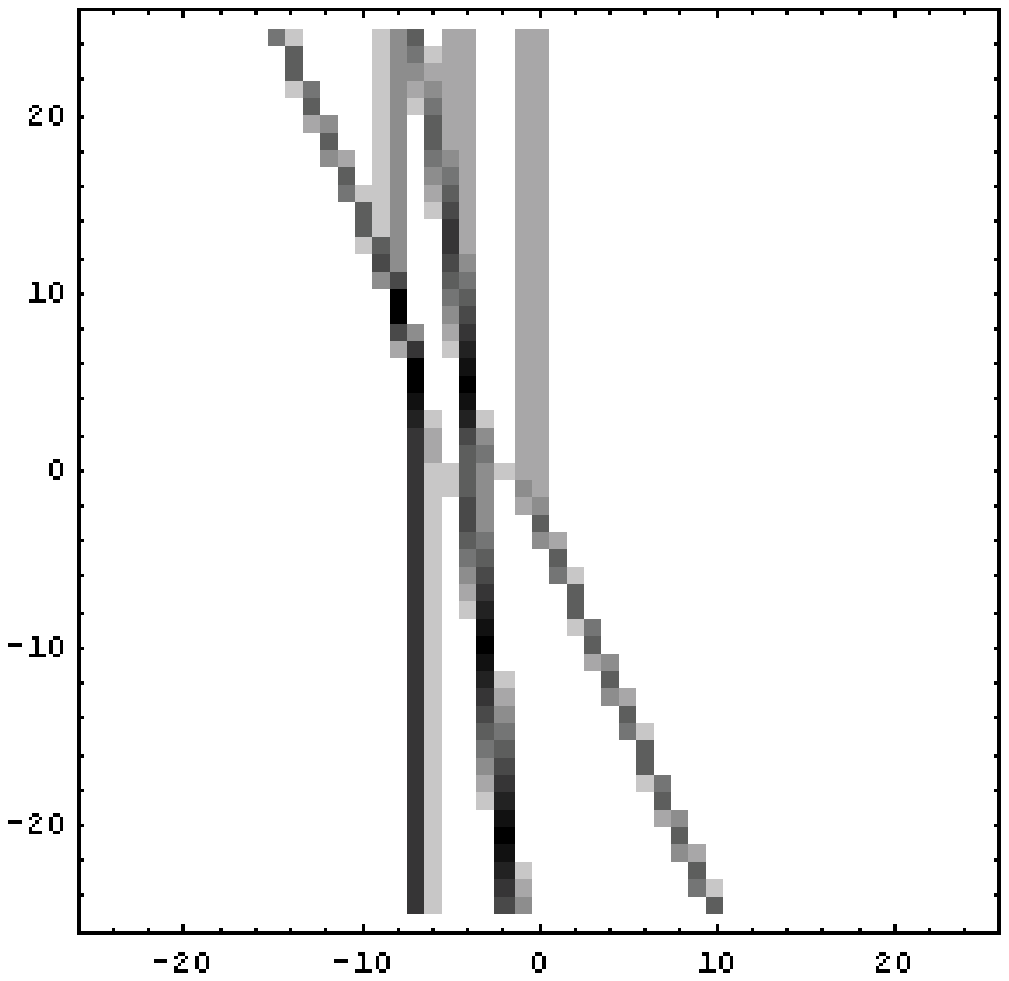}}
\kern-0.385\textwidth
\hbox to \textwidth{\hss(c)\kern0em\hss(d)\kern4em}
\kern+0.355\textwidth
\caption{Snapshots illustrating the temporal evolution 
of a $(3,3)$-resonant soliton solution for Eq.~\eqref{rutoda}, with
$P_1=-10$, $P_2=-7$, $P_3=-5$, $P_4=-1$, $P_5=4$, $P_6=5$, 
$\r= 7$, $\s= 4$, $A_1=-8$, $A_2=-6$, $A_3=0$, $A_4=2$, $A_5=4$, $A_6=7$: 
(a) $l=-15$, (b) $l=-10$, (c) $l=0$, (c) $l=10$.}
\label{(3,3)}
\end{figure}
%%%%%%%%%%%%%%%%%%%%%%%%%%%%%%%%%%%%%%%%%%%%%%%%%%%%%%%%%%%%%%%%%%%%%%%%%

%%%%%%%%%%%%%%%%%%%%%%%%%%%%%%%%%%%%%%%%%%%%%%%%%%%%%%%%%%%%%%%%%%%%%%%%%
\begin{figure}[t!]
%\begin{center}
%\subfigure[$l=-10$]{\includegraphics[scale=0.5]{4x4-time-10.eps}}
%\subfigure[$l=0$]{\includegraphics[scale=0.5]{4x4-time0.eps}}
%\subfigure[$l=10$]{\includegraphics[scale=0.5]{4x4-time10.eps}}
%\subfigure[$l=15$]{\includegraphics[scale=0.5]{4x4-time20.eps}}
%\end{center}
\centerline{
\includegraphics[scale=0.5]{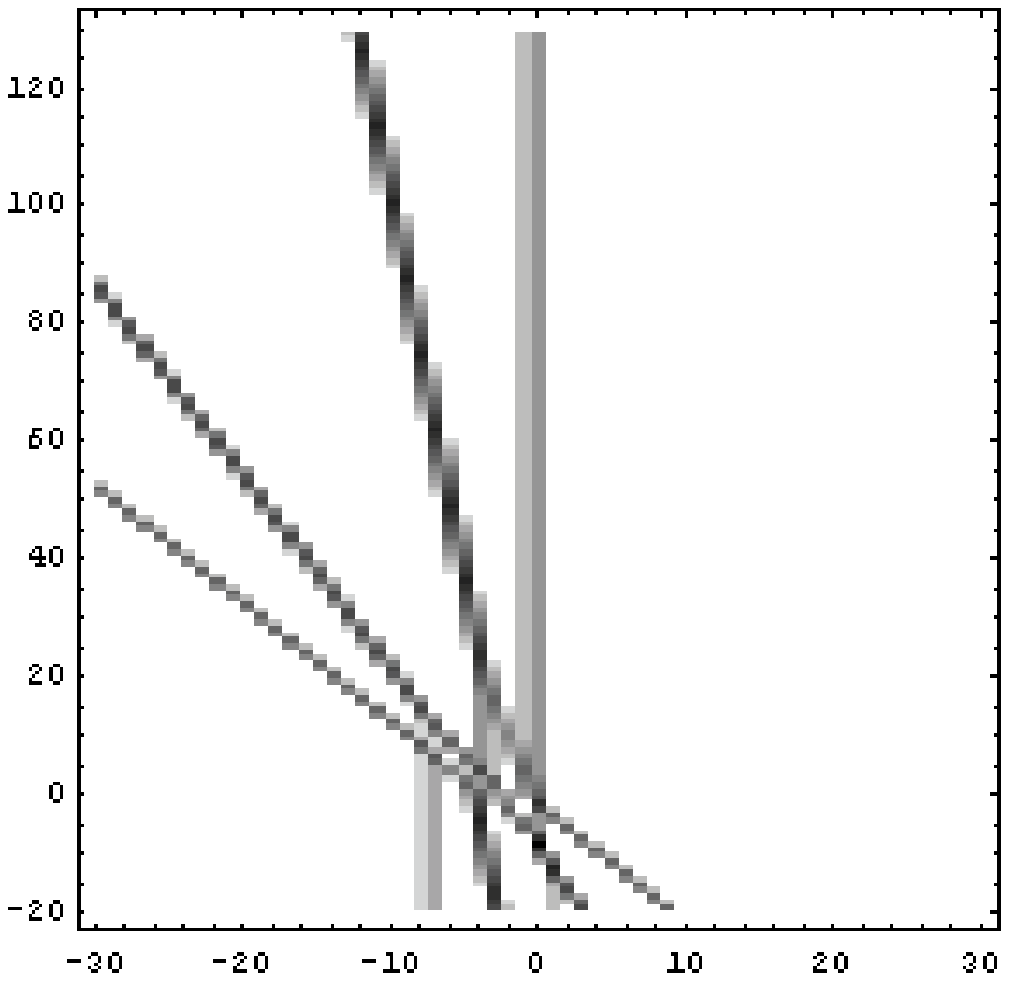}\quad\ 
\includegraphics[scale=0.5]{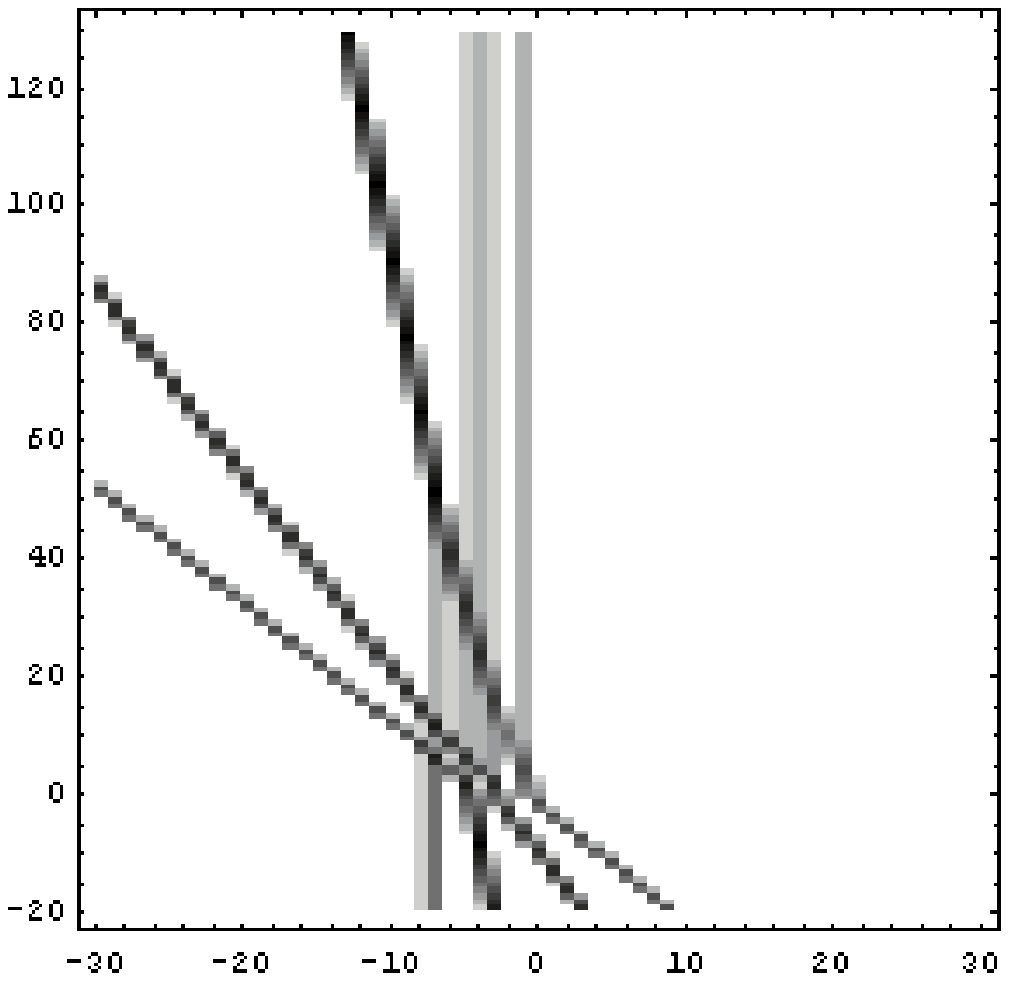}}
\kern-0.385\textwidth
\hbox to \textwidth{\hss(a)\kern0em\hss(b)\kern4em}
\kern+0.345\textwidth
\vglue\bigskipamount
\centerline{
\includegraphics[scale=0.5]{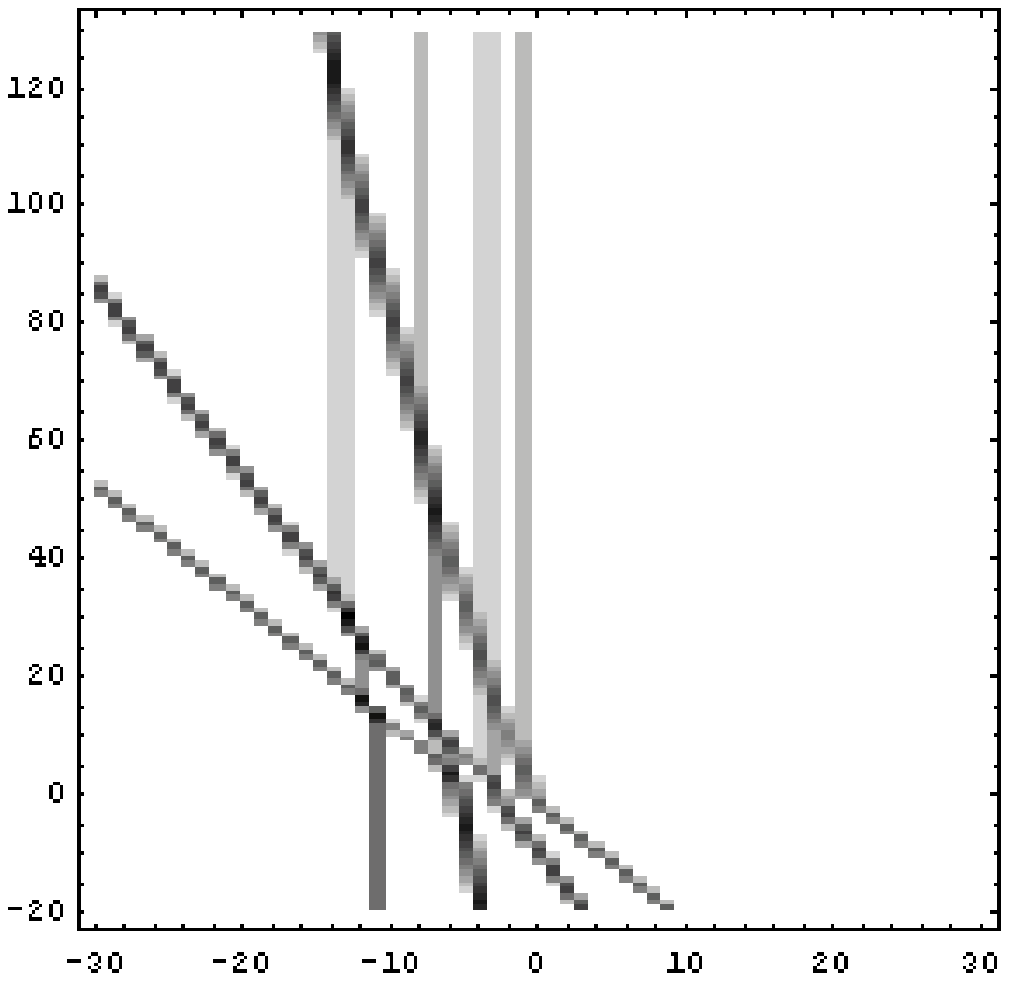}\quad\ 
\includegraphics[scale=0.5]{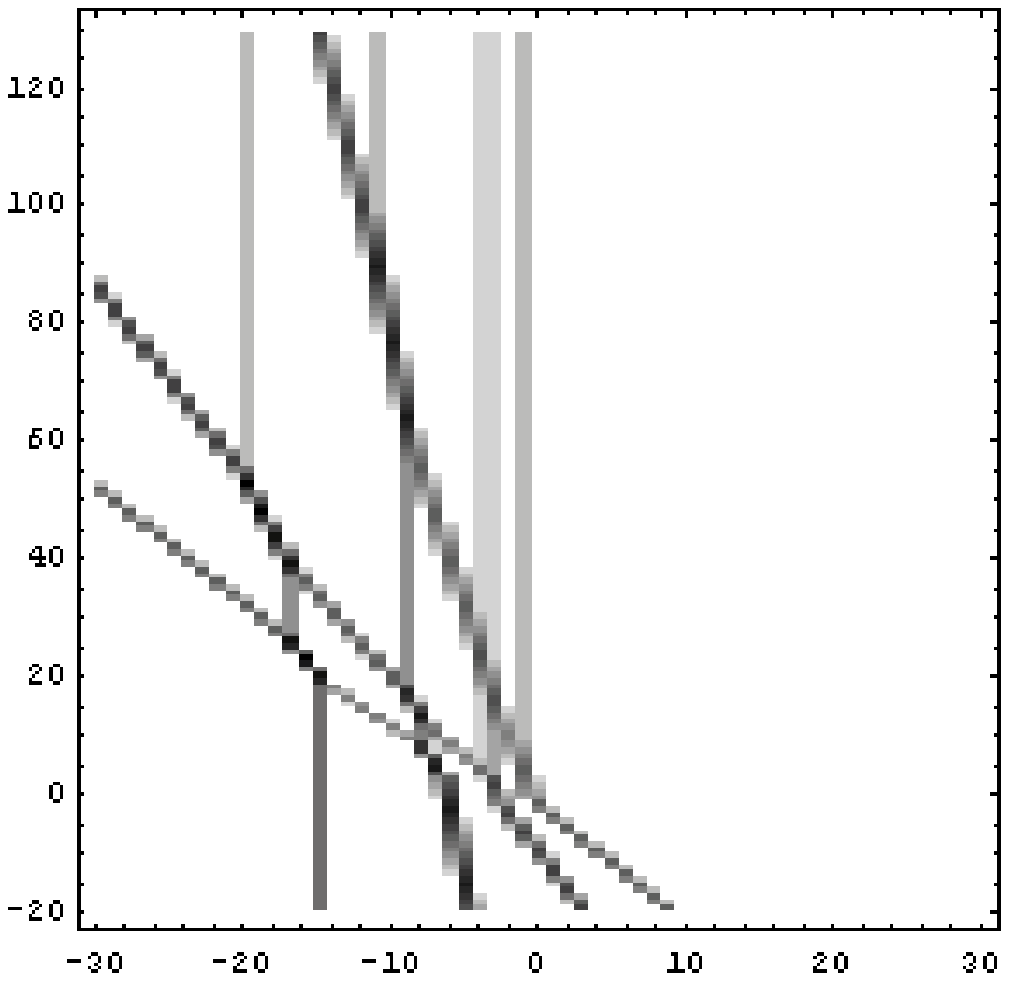}}
\kern-0.385\textwidth
\hbox to \textwidth{\hss(c)\kern0em\hss(d)\kern4em}
\kern+0.355\textwidth
\caption{Snapshots illustrating the temporal evolution of a 
$(4,4)$-resonant soliton solution for Eq.~\eqref{rutoda}, with
$P_1=-15$, $P_2=-12$, $P_3=-9$, $P_4=-3$, $P_5=1$, $P_6=1$, $P_7=4$, $P_8=7$,
$\r= 7$, $\s= 4$, 
$A_1=-8$, $A_2=-6$, $A_3=0$, $A_4=2$, $A_5=4$, $A_6=7$, $A_7=8$, $A_8=10$:
(a) $l=-10$, (b) $l=0$, (c) $l=10$, (d) $l=20$.}
\label{(4,4)}
\end{figure}
%%%%%%%%%%%%%%%%%%%%%%%%%%%%%%%%%%%%%%%%%%%%%%%%%%%%%%%%%%%%%%%%%%%%%%%%%

Like in the 2D Toda lattice~\eqref{2dTL}
and its fully discrete version~\eqref{e:discrete2dToda},
we now consider more general resonant solutions for the ultra-discrete
2DTL~\eqref{rutoda}.
We start from the general $\tau$~function defined in 
Eq.~\eqref{tau3}, and
introduce again the parameters
$\e^{P_k/\epsilon} = p_k$ and
$e^{A_k/\epsilon} = \alpha_k$
($k=1, 2,\dots ,\M$) and the variable
$\rho_{l,m,n}^\epsilon = \epsilon \log \tau_{l,m,n}$,
together with $\e^{-\r/\epsilon} = \delta$ and
$\e^{-\s/\epsilon} = \kappa$.
Taking the limit $\epsilon \to 0^+$, we then obtain
the following solution of
the ultra-discrete 2DTL~\eqref{rutoda}: 
\begin{equation}
\rho_{l,m,n}
 =\max_{1\leq i_1<\cdots < i_\N\leq \M}
\Bigg[\sum_{j=1}^\N R_{i_j}+2\sum_{j=2}^\N(j-1)P_{i_j}\Bigg]\,,
\label{e:tauplus}
\end{equation}
where again $\lim_{\epsilon\to 0^+} \rho_{l,m,n}^\epsilon= \rho_{l,m,n}$,
with the maximum being taken among all possible combinations of 
the indices~$i_j$ $(j=1,\dots,\N)$,
and where once more we have
\begin{eqnarray}
R_i = A_i+ n P_i + l \max(0,P_i-\r)-m \max(0,-P_i-\s)\,.
\end{eqnarray}
%%%%\nonumber \\
%%%%&&\quad \quad -
%%%%2\sum_{k=\M-\Np+2}^{\M}(k-1)P_{k}
%%%%+\sum_{k=\M-\Np+1}^{\M}
%%%%(A_{k}+nP_{k}+t\max(0,P_{k}-\r)-m\max(0,-P_{k}-\s))
%%%%)\nonumber\\

Equation~\eqref{e:tauplus} produces complicated soliton solutions displaying 
resonance and web structure.
As an example, in Fig.~\ref{(3,3)} and Fig.~\ref{(4,4)}
we show some snapshots of the time evolution of a
$(3,3)$-resonant soliton solution and 
a $(4,4)$-resonant soliton solution. 
Indeed, we conjecture that, 
similarly to its counterparts for the 2DTL and in its fully discrete analogue,
Eq.~\eqref{e:tauplus} yields the $(\Np,\Nm)$-soliton solution of the 
ultra-discrete~2DTL Eq.~\eqref{rutoda}, with $\Nm=\N$ and $\Np=\M-\N$. 
Unlike the semi-continuous and fully discrete cases, however, 
we were unable to prove this conjecture using the techniques
introduced in Ref.~\cite{JPhysA2003v36p10519}.
In this respect, it should be noted that solutions of the 
ultra-discrete~2DTL arise as a result of the properties of the 
maximum function, and therefore their study might require 
the use of techniques from tropical algebraic geometry, 
which is a subject of current research~\cite{RichterGebert,Sturmfels,Speyer}.

It should also be noted that the interaction patterns in the 
ultra-discrete system differ somewhat from their analogues 
in the semi-continuous and fully discrete cases.
In particular, low-amplitude interaction arms may disappear
when considering the ultra-discrete limit.
Furthermore, the specific interaction patterns %which are produced
in the ultra-discrete limit
depend on the value of the parameters $r$ and $s$,
and different kinds of solutions may appear for different values
of~$r$ and~$s$.
In particular, large values of $r$ and~$s$ tend to result in the
production of several vertical solitons,
as shown in Figs.~\ref{(3,3)} and~\ref{(4,4)}.
In order to preserve the soliton count in these cases,
all the outgoing vertical solitons should be counted as one,
as should the incoming ones. 
In this sense, a set of outgoing or incoming vertical lines can be 
considered as a bound state of several solitons.
A full characterization of these phenomena and their parameter
dependence is however outside the scope of this work.
%and is a subject for future research.

\section{Conclusions}

We have demonstrated the existence of soliton resonance and web structure
in discrete soliton systems by presenting a class of solutions of 
the two-dimensional Toda lattice (2DTL) equation, its fully discrete
analogue and their ultra-discrete limit.
Soliton resonance and web structure had been previously found
for nonlinear partial differential equations such as the 
KP and cKP systems.
Note that the 2DTL is a differential-difference equation,
its fully discrete version is a difference equation and
their ultra-discrete limit is a cellular automaton, therefore,
our findings show that resonance and web structure phenomena 
are rather general features of two-dimensional integrable systems
whose solutions are expressed in determinant form.

A full characterization of the solutions,
including the study of asymptotic amplitudes and velocities
and the resonance condition was provided both in the semi-continuous 
and in the fully discrete case.
Their analogue in the ultra-discrete case, together with an
analysis of the intermediate patterns of interactions
is outside the scope of this work, and remains as a problem 
for further research.
Of particular interest is the ultra-discrete 2DTL,
where new types of solutions such as the L-shape soliton 
shown in Fig.~\ref{V-soliton} appear.

Finally, we note that the class of solutions presented 
in this work is just one of the possible choices that yield resonance 
and web structure.  
Just like with the KP and cKP equations, the class of soliton solutions 
of each of the systems we have considered (namely, the 2DTL and its 
fully discrete and its ultra-discrete analogues) is much wider, 
and includes also partially resonant solutions.  
The solutions described in this work represent the extreme case 
in which all of the interactions among the various solitons are resonant, 
whereas ordinary soliton solutions represent the opposite case where 
none of the interactions among the solitons are resonant.
Inbetween these two situations, a number of intermediate cases
exist in which only some of the interactions are resonant.
As in the case of the KP equation, the study of these 
partially resonant solutions remains an open problem.

\section*{Acknowledgements}

We thank M.~J.\ Ablowitz, S. Chakravarty, Y. Kodama, J. Matsukidaira and A. Nagai for 
many helpful discussions.  
K.M.\ acknowledges support from the Rotary foundation and the 21st Century 
COE program ``Development of Dynamic Mathematics with High Functionality'' 
at Faculty of Mathematics, Kyushu University. 
G.B.\ was partially supported by the National Science
Foundation, under grant number DMS-0101476.

\end{document}